\documentclass[a4paper]{scrartcl}
\usepackage{ifthen}
\usepackage{epsfig}
\usepackage{amssymb}
\usepackage{graphicx}
\usepackage{amsmath}
\usepackage{array}
\usepackage[center]{subfigure}

\newcommand{\ba}{\begin{eqnarray}}
\newcommand{\ea}{\end{eqnarray}}
\newcommand{\be}{\begin{equation}}
\newcommand{\ee}{\end{equation}}
\newcommand{\DS}[1]{/\!\!\!#1}

\makeatletter
\def\fmslash{\@ifnextchar[{\fmsl@sh}{\fmsl@sh[0mu]}}
\def\fmsl@sh[#1]#2{%
  \mathchoice
    {\@fmsl@sh\displaystyle{#1}{#2}}%
    {\@fmsl@sh\textstyle{#1}{#2}}%
    {\@fmsl@sh\scriptstyle{#1}{#2}}%
    {\@fmsl@sh\scriptscriptstyle{#1}{#2}}}
\def\@fmsl@sh#1#2#3{\m@th\ooalign{$\hfil#1\mkern#2/\hfil$\crcr$#1#3$}}
\makeatother

\begin{document}
\begin{titlepage}\begin{flushright}
SI-HEP-2009-06\\
LPT-09-60
\vspace{0.6cm}
\end{flushright}
\vfill
\begin{center}
{\Large\bf 
Semileptonic charm decays $D\!\to\! \pi  \ell \nu_\ell$ and $D\!\to\! K \ell \nu_\ell$ 
from QCD Light-Cone Sum Rules}\\[2cm]
{\large\bf  
A.~Khodjamirian\,$^{(a)}$, 
Ch.~Klein\,$^{(a)}$, Th.~Mannel\,$^{(a)}$ and N.~Offen\,$^{(b)}$}\\[0.5cm]
{\it  $^{(a)}$\,Theoretische Physik 1, Fachbereich Physik,
Universit\"at Siegen,\\ D-57068 Siegen, Germany }\\[2mm]
{\it  $^{(b)}$\,Laboratoire de Physique Theorique CNRS/Univ.
Paris-Sud 11,\\ F-91405 Orsay, France }
\end{center}
\vfill
\begin{abstract}
We present a new calculation of the $D\to\pi$ and $D \to K$ 
form factors from QCD light-cone sum rules.  
The $\overline{MS}$ scheme for the $c$-quark mass is used 
and the  input parameters are updated. 
The results are $f^+_{D\pi}(0)= 0.67^{+0.10}_{-0.07}$,  
$f^+_{DK}(0)=0.75^{+0.11}_{-0.08}$ and 
$f^+_{D\pi}(0)/f^+_{DK}(0)=0.88 \pm 0.05$. Combining
the calculated form factors  with the latest CLEO data,
we obtain 
$|V_{cd}|=0.225\pm 0.005 \pm 0.003\ ^{+0.016}_{-0.012}$
and $|V_{cd}|/|V_{cs}|= 0.236\pm 0.006\pm 0.003\pm 0.013$
where the first and second errors are of experimental 
origin and the third error is due to the estimated uncertainties
of our calculation. We also evaluate
the form factors $f^-_{D\pi}$ and $f^-_{DK}$ and predict 
the slope parameters  at $q^2=0$. 
Furthermore, calculating the form factors from 
the sum rules at $q^2<0$, we fit them to various parameterizations. 
After analytic continuation, the  
shape of the $D\to \pi,K $ form factors in the whole 
semileptonic region is reproduced, in a good agreement 
with experiment. 
\end{abstract}
\vfill
\end{titlepage}

\section{Introduction}

Recent measurements of the semileptonic 
$D\to \pi \ell \nu_\ell$  and $D\to K \ell \nu_\ell$  decays 
by CLEO collaboration \cite{CLEO1,CLEO2} 
provide new accurate results on branching fractions 
and differential decay rates, in addition to   
the previously accumulated  data \cite{oldCLEO,FOCUS,BELLE,BABAR}.
The decay rate distributions in the bins of the variable $q^2$
(the invariant mass squared of the lepton pair), 
yield the products of transition form factors  
and CKM matrix elements,
$|V_{cd}|f^+_{D\pi}(q^2)$ and $|V_{cs}|f^+_{DK}(q^2)$. In addition, 
the form factor shapes are reconstructed and fitted  
to various parameterizations.
With the new data available, it is timely to update the 
theoretical analysis of the $D\to \pi$ and 
$D\to K$ form factors, aiming at more accurate determination
of $|V_{cd}|$ and  $|V_{cs}|$.

In this paper, we  recalculate the  $D\to \pi,K$ form factors 
from QCD light-cone sum rules (LCSR's). 
In this method \cite{lcsr}, the correlation 
function of quark currents is constructed in a form of 
a transition matrix element 
between the vacuum and the final hadron state. 
In our case, the quark current with $D$-meson quantum numbers
is correlated  with the charmed weak current, whereas the hadron state is the on-shell pion or  kaon. Two different 
representations of  the correlation function are then 
equated:
the operator-product expansion (OPE) near the light-cone
and the dispersion integral over hadronic states. 
In the latter, the ground $D$-meson  
contribution containing the $D\to \pi$ or $D\to K$ form factor is 
singled out.  Applying the quark-hadron duality approximation, 
the remaining dispersion integral over the 
higher states is replaced by the integral over the OPE spectral density. 
The LCSR approach, though having a limited accuracy,
provides analytical expressions for the form factors,
in terms of finite quark masses 
and universal light-cone  distribution amplitudes (DA's) 
of pion or kaon.  Importantly,  
the  heavy-light form factors calculated from 
LCSR's with gluon radiative corrections, include both 
``hard-scattering''  and ``soft-overlap'' components, and  
the latter is predicted to be the dominant one. 

The fact that the correlation function
is calculated at a finite heavy quark mass,
simplifies our task, because 
the OPE of  $b$-quark and $c$-quark currents have the same 
analytical expressions. Only the quark mass value has to be 
changed and the normalization scales have to be adjusted.
Hence, for example, the LCSR for  
$f^+_{D\pi} $  represents a by-product 
of the LCSR obtained for $f^+_{B\pi}$ . 
The latter form factor
used  to determine  $|V_{ub}|$ is the most familiar application \cite{Bpilcsr,BBKR,BpilcsrNLO,BBB,BZ04,DKMMO} 
of this method
\footnote{Interestingly, one of the earliest applications 
\cite{BBD} of the ``sum rules on the light-cone'' 
was to the $D\to K^{(*)}$ form factors.}. In what follows, we employ 
the recent update of LCSR for $B\to \pi$ form factors
presented  in \cite{DKMMO}. 
Importantly, the $D\to \pi,K$ form factors obtained from the sum rules,
being confronted with experimental data, provide an important 
test of the whole method.

Compared with the previous calculations of $D\to\pi,K$ form factors
from LCSR's \cite{KRWWY,BallDpi}, certain 
modifications and improvements are made. First of all, following \cite{DKMMO}, we systematically 
use the $\overline{MS}$ scheme for  the $c$-quark mass, 
whereas earlier calculations switched 
to the pole mass in the final sum rules. 
In this respect we benefit from recent accurate 
determinations \cite{mc} of the $c$-quark mass 
from the charmonium QCD sum rules. Also 
the strange quark mass entering the sum rule for $D\to K$ 
form factor
has a smaller uncertainty than before. 
Furthermore, we use the improved  determination \cite{CKP} 
of the $SU(3)_{fl}$ violating Gegenbauer 
coefficient of the twist-2 kaon DA, and 
the updates of the pion 
and kaon twist-3,4 DA's from \cite{BBL}.

The main novelty in this paper concerns
the $q^2$-dependence   of the form factors.
First, we determine the slope parameters
at $q^2=0$, which involve the second form factor $f^-_{D\pi(K)}$. 
The latter is calculated using the same method and input.
Note that for $D\to \pi(K)$ transitions, LCSR's 
are applicable only in the lower part of the region 
$0<q^2<(m_D-m_{\pi(K)})^2$, accessible in the semileptonic 
$D\to \pi(K) \ell \nu_\ell$ decays. At $q^2$ 
approaching  $m_c^2$, the virtuality of the $c$-quark 
becomes a soft scale, and OPE is not reliable. 
In this paper we predict the form factor shapes,
combining  LCSR with analyticity.  
We employ the spacelike region $q^2<0$, where the light-cone OPE  works even better, than at small positive $q^2$.  
The LCSR results for the form factors  at $q^2<0$ 
are fitted to various parameterizations, such as 
the dispersion relation with an effective pole \cite{BK}  and 
the recent version of series parameterization \cite{BCL}. 
We then make use of  
the analytic continuation from negative to positive $q^2$ 
and predict the form factors in the whole  semileptonic region. 

The paper is organized as follows. 
In Section 2 we present the
outline of the method and discuss the expected accuracy 
of our calculation. In Section 3 the input parameters
are listed and in Section 4 the numerical results for 
the form factors at zero momentum transfer are presented. 
In Section 5 we compare the calculated form factors 
with experimental data, 
and determine $|V_{cd}|$ and $|V_{cs}|$.  
In Section 6 we turn to the determination of the form factor 
shape and present the fit results for various parameterizations.
The calculated shapes are used to predict the total
semileptonic widths.
Finally, Section 7 contains a concluding discussion.
In Appendix A the relevant definitions of light-cone DA's
and their expressions used in LCSR's are collected, and 
in Appendix B the formulae for the contributions to LCSR's
are presented.

\section{Outline of the LCSR method}

The central objects of our interest are the $D\to \pi$ 
form factors $f^{\pm}_{D\pi}(q^2)$
defined in a standard way  from the hadronic matrix element: 
\be
\langle\pi^-(p)\mid \bar{d}\gamma_\mu c\mid D^0(p+q)\rangle
=2f^+_{D\pi}(q^2) p_\mu +(f^+_{D\pi}(q^2)+f^-_{D\pi}(q^2)) q_\mu ~,
\label{def}
\ee
and the analogous form factors $f^{\pm}_{DK}(q^2)$ 
of $D^0\to K^-$ transition, obtained 
by replacing $d\to s$ and $\pi^-\to K^-$ in the above.
In what follows, we work in the isospin symmetry limit,
so that the $D^+\to K^0$  and $D^0\to K^-$ hadronic matrix elements are equal
and  the $D^+\to \pi^0$ form factors are obtained by multiplying $f^\pm_{D\pi}(q^2)$ with 
$1/\sqrt{2}$.
 The form factor $f^-$ is combined with   
$f^+$ in the scalar form factor 
 \be
f^0_{D\pi(K)}(q^2) = f^+_{D\pi(K)}(q^2) + \frac{q^2}{m_D^2-m_{\pi(K)}^2} f^-_{D\pi(K)}(q^2) \,,
\label{eq:f0}
\ee 
which plays a minor role in semileptonic transitions and is 
``visible'' only  in $D\to K \mu \nu_\mu$.

The correlation function used to derive LCSR's 
for $D\to \pi$ form factors is defined as  
\ba
F_\mu^{\pi}(p,q) & = & i \int d^4x e^{i q \cdot x} \langle \pi(p) |
T \left\{ \bar{d}(x) \gamma_\mu c(x), m_c \bar{c}(0) i\gamma_5 u(0)
\right\} | 0 \rangle
\nonumber \\
&=& F^{\pi}(q^2,(p+q)^2)p_\mu +\widetilde{F}^{\pi}(q^2,(p+q)^2)q_\mu\,
\label{eq:corr}\,.
\ea 
Replacing $d\to s$ and $\pi\to K$ in the above,
we obtain $F_\mu^{K}$, the  correlation function
for $D\to K$ form factors. 
The two invariant amplitudes $F^{\pi(K)}$  and 
$\widetilde{F}^{\pi(K)}$ yield 
two separate sum rules for $f^+_{D\pi(K)}$ 
and $(f^+_{D\pi(K)}+f^-_{D\pi(K)})$, respectively. Using 
(\ref{eq:f0}), one then obtains $f^0_{D\pi(K)}$, so that 
in our calculation, by default,  $f^0_{D\pi(K)}(0)=f^+_{D\pi(K)}(0)$.

At $q^2 \ll m_c^2$ and $(p+q)^2\ll m_c^2$, 
the $c$-quark propagating in the correlation function 
has a large virtuality and the product of the $c$-quark 
fields is expanded near the light-cone $x^2\sim 0$. 
This expansion starts in leading order (LO) from the free 
$c$-quark propagator and includes the $O(\alpha_s)$ (NLO) corrections due to hard-gluon exchanges between the quark lines 
and soft-gluon emission. 
 A more detailed derivation can be found e.g. in \cite{BBKR},
where also the origin of the light-cone expansion 
in the correlation function is explained.

The OPE result for the correlation function (\ref{eq:corr}) 
is cast in  a factorized form, where the perturbatively calculable 
kernels  are convoluted with the pion DA's of growing twist
$t=2,3,4,..$. For the invariant amplitude $F^\pi$ one obtains:
\be
\Big[F^\pi(q^2, (p+q)^2)\Big]_{OPE}=\sum\limits_{t=2,3,4}
F^{\pi,t}_0(q^2,(p+q)^2)+
\frac{\alpha_sC_F}{4\pi}\sum\limits_{t=2,3}
F^{\pi,t}_1(q^2,(p+q)^2)\,,
\label{eq:exp}
\ee
where the LO\,(NLO) parts in $\alpha_s$ are 
the convolutions 
\be
F^{\pi,t}_{0(1)}(q^2, (p+q)^2)=\int {\cal D} u~
 T^{(t)}_{0(1)}(q^2,(p\!+\!q)^2,m_c^2,u,\mu)\, 
\phi^{(t)}_\pi(u,\mu)\,.
\label{eq:conv}
\ee 

The perturbative kernels $T^{(t)}_{0,1}$ stem from 
the $c$-quark propagator, and $T^{(t)}_{1}$ 
include the loops with hard-gluon  exchanges in $O(\alpha_s)$.
The pion DA's $\phi^{(t)}_\pi(u,\mu)$ 
represent universal vacuum-pion matrix elements of  
light-quark and gluon operators. The simplest bilocal 
quark-antiquark operators $\bar{d}(x)\Gamma^a u(0)$ 
(where $\Gamma^a$  is a generic combination of $\gamma$-matrices) originate after contracting the free $c$-quark fields in 
the correlation function (\ref{eq:corr}) .
In addition, soft gluons emitted  from the propagating 
$c$-quark, together with light quarks and antiquarks, form
DA's of higher multiplicity, starting from 
the three-particle (quark-antiquark-gluon) DA's of $t=3,4$.  
In (\ref{eq:conv}), the integration over $u$ is a 
generic notation for the momentum distribution 
between  constituents of the pion in DA's. 
The definitions and explicit expressions for all 
relevant DA's are given in  Appendix A. 
The presence of the terms with $2\leq t\leq 4$ in 
(\ref{eq:exp}) reflects 
the currently achieved accuracy in OPE: the twist 2,3,4 terms in LO
(including three-particle contributions of twist 3,4) and 
the twist-2 and twist-3\,(two-particle DA's) terms in NLO.

The factorization scale $\mu$ in (\ref{eq:conv}) separates
the perturbative kernels dominated by near light-cone distances 
from the long-distance quark-gluon dynamics  
in DA's. The collinear
divergences in the twist-2 of OPE are absorbed 
in the logarithmic evolution of DA's
as shown in \cite{BpilcsrNLO,BBB}. 
The factorization for the twist-3 part is proved  
for the asymptotic DA's in \cite{BZ04} 
and confirmed in \cite{DKMMO}.

Note that the two terms in the $c$-quark propagator 
proportional to $m_c$ and $\DS p_c$ yield 
two different parts of OPE with even ($t=2,4,..$)
and odd ($t=3,5,..$) twists,  
respectively, corresponding to different chiralities of
$\Gamma^a$  matrices in the light-quark operators. Hence, the twist expansion goes in even and odd
twists separately.    
The two most important LO contributions of twist-2 and twist-3 
two-particle DA's to the correlation function 
originate from the $\sim m_c$ 
and $\sim \DS p_c$ parts
of the free $c$-quark propagator, respectively, and have 
simple  expressions:
\ba
F^{\pi,2}_{0}(q^2,(p+q)^2)&=& 
f_\pi m_c^2\int\limits_0^1\frac{du\,\,\varphi_\pi(u)}{m_c^2-q^2\bar{u}-
(p+q)^2u}\,,
\nonumber\\
F^{\pi,3 [two\, part.]}_{0}(q^2,(p+q)^2)&=&f_\pi\mu_\pi m_c \int\limits_0^1\frac{du\,\,
}{m_c^2-q^2\bar{u}-(p+q)^2u}
\Big\{\phi_{3\pi}^p(u)
\nonumber\\
&+& \frac{1}{6}\Big[2+\frac{m_c^2+q^2}{m_c^2-q^2\bar{u}-(p+q)^2u}\Big]
\phi_{3\pi}^\sigma(u)
\Big\}\,,
\label{eq:LO}
\ea
where $\mu_{\pi}=m_\pi^2/(m_u+m_d)$ and $\bar{u}=1-u$.  
In the above, $\varphi_\pi$ and  
$\phi_{3\pi}^p$, $\phi_{3\pi}^\sigma$ are the pion twist-2 and twist-3 two-particle DA's, respectively. Note that  
the formal $1/m_c$ suppression of $F^{\pi,3}_{0}$ versus $F^{\pi,2}_{0}$  
is overwhelmed numerically,  because the enhanced  
light-quark parameter $\mu_{\pi} $ is  
larger than $m_c$.
The remaining twist-3 term in LO,  
due to the three-particle DA, not shown in (\ref{eq:LO}), is 
strongly suppressed by the ratio of the small normalization factor 
$f_{3\pi}$ to $m_c$.  

Turning to higher-twist ($t>3$) contributions, 
one has to take into 
account  that in the light-cone OPE  each two units of twist are 
accompanied by an extra $x^2$, yielding 
an additional power of the denominator 
\be
D=\frac{1}{ m_c^2-q^2\bar{u}-(p+q)^2u}, \hspace{0.5cm}
(0\leq u \leq 1)\,,
\label{eq:D}
\ee
 in the correlation
 function. Hence, for example,
 the contribution $F_{0}^{\pi,4}$ of the twist-4  
two- and three-particle DA's is subleading 
with respect to the twist-2 part $F_{0}^{\pi,2}$ in (\ref{eq:LO}), with a 
suppression factor $\sim \delta_\pi^2 \langle D\rangle$ , 
where $\langle D\rangle$ is the weighted (over DA's)  average 
of the denominator ($\ref{eq:D}$),
and $\delta_\pi^2\sim \Lambda_{QCD}^2$ is the normalization factor of the twist-4 DA's. 
 Similarly, the twist-5 contributions, not yet included 
in the currently  used version (\ref{eq:exp}) of  OPE,
are expected to be suppressed with respect to the twist-3 terms,   
parametrically $\sim \Lambda_{QCD}^2\langle D\rangle$. 
We emphasize that the suppression of higher-twist 
contributions is effective only  if   
both external momenta  squared 
$q^2$ and $(p+q)^2$ in (\ref{eq:D}) are kept $\ll m_c^2$, that is,
when  the $c$-quark is sufficiently virtual. 

The detailed calculation of the correlation 
function, including $O(\alpha_s)$ corrections to 
the twist-2 and twist-3 part, is given in previous papers 
and we will not repeat it. Here we use the recent 
update in terms of $\overline{MS}$  heavy quark mass 
presented in \cite{DKMMO}, where $m_b$ has to be replaced
by $m_c$, with a corresponding adjustment of the scales.
As in \cite{DKMMO}, we use a universal  normalization 
scale $\mu$ for $\alpha_s$, quark masses  
and DA's . 
The $u$-, $d$-quark masses and $m_\pi^2$  
are neglected  everywhere, except in the 
parameter $\mu_\pi$. 

The correlation function $F^K_\mu $ for the $D\to K$ form factors
includes $SU(3)_{fl}$ violation effects of different origin, starting from $O(m_s)\sim O(m_K^2)$. One effect is purely kinematical, due to 
the presence of $p^2=m_K^2\neq 0$ 
in the $c$-quark propagator. The expressions 
for the perturbative kernels have to be modified, so that the term 
$m_K^2u\bar{u}$ has to be added to the denominator (\ref{eq:D}). 
Furthermore, there are $SU(3)_{fl}$ violating
corrections  in DA's. In twist 2, the deviations of 
$\varphi_K$ from $\varphi_\pi$ is due to
differences between the normalization
parameters ($f_K$ vs $f_\pi$) and Gegenbauer moments. Importantly,
one has to take into account the first Gegenbauer moment of the kaon DA, 
$a_1^K\neq$ 0, responsible for the momentum distribution 
asymmetry between the strange quark and nonstrange antiquark in the kaon. 
In general, also $a_2^K \neq a_2^\pi$. In twist 3 and 4  DA's 
one has to replace the normalization parameters: $\mu_\pi\to 
\mu_K=m_K^2/(m_u+m_s)$, $f_{3\pi}\to f_{3K}$ and $\delta^2_\pi\to \delta^2_K$,
respectively. In addition, there 
are $O(m_K^2)$ admixtures of the twist-2 DA in the kaon  
DA's  of twist 3 and 4. All these and some other less 
important $SU(3)_{fl}$ violating effects are included  in the expressions 
for the kaon DA's \cite{BBL} 
presented in Appendix A.

To access the 
$D\to \pi, K $  form factors, one equates  the OPE result 
to the hadronic dispersion
relation for the correlation function in the variable 
$(p+q)^2$, the momentum squared  
of the $D$-meson interpolating current.
In the $D\to \pi$  case, the resulting relation for the 
invariant amplitude $F^\pi$ is:
\be
\Big[F^\pi(q^2,(p+q)^2)\Big]_{OPE}= 
\frac{2m_D^2f_D  f_{D\pi}^+(q^2)}{m_D^2-(p+q)^2}+
\frac{1}{\pi}\int\limits_{s_0^D}^{\infty}ds 
\frac{\Big[\mbox{Im}F^\pi(q^2,s)\Big]_{OPE}}{s-(p+q)^2}\,.
\label{eq:SRraw}
\ee
In the above, the $D$-meson pole
term (with the mass $m_D$ and the decay constant defined as 
$\langle 0 \!\! \mid \! m_c\bar{q}i\gamma_5 c \!\mid\! \! D\rangle =m_D^2f_D$), contains the desired form factor $f_{D\pi}^+$.  Also
in (\ref{eq:SRraw})
the quark-hadron duality approximation is applied, replacing  
the hadronic spectral density of the higher states by the OPE  spectral density. 
The latter approximation introduces
the threshold parameter $s_0^D$ which is non-universal 
and has to be determined for each sum rule independently.   
Note that $s_0^D$ is an effective parameter, 
not necessarily equal to the lowest hadronic continuum threshold 
$(m_{D^*}+m_\pi)^2$. 

After subtracting the integral on r.h.s. of 
(\ref{eq:SRraw}) from both sides of this equation, 
one performs  Borel transformation of (\ref{eq:SRraw}) 
replacing the variable $(p+q)^2$ with the Borel parameter 
$M^2$ and exponentiating the denominators. 
E.g., the powers of the denominator $D$ in (\ref{eq:D}) transform
as:
\be
{\cal B}_{(p+q)^2\to M^2} \big\{D^n\big\}=
\frac{1}{(n-1)!u^n(M^2)^{n-1}}
e^{-\frac{m_c^2-q^2\bar{u}}{uM^2}}\,.
\label{eq:Borel}
\ee
Finally, the LCSR for the $D\to \pi$ form factor is obtained:
\ba
f^+_{D\pi}( q^2)=\frac{e^{m_D^2/M^2}}{2m_D^2f_D}
\Bigg ( \sum\limits_{t=2,3,4}F_{0}^{\pi,t}(q^2,M^2,s_0^D)
+\frac{\alpha_sC_F}{4\pi} \sum\limits_{t=2,3}F_{1}^{\pi,t}(q^2,M^2,s_0^D)
\Bigg )~,
\label{eq:LCSR}
\ea
where the functions $F_{0}^{\pi,t}(q^2,M^2,s_0^D) $ and 
$F_{1}^{\pi,t}(q^2,M^2,s_0^D)$ 
are derived applying the Borel-and-subtraction  
procedure to each twist component of the OPE in (\ref{eq:exp}):
\be
F_{0(1)}^{\pi,t}(q^2,M^2,s_0^D)= 
{\cal B}_{(p+q)^2\to M^2}
\Big\{F_{0(1)}^{\pi,t}(q^2,(p+q)^2)\Big \}-
\frac{1}{\pi}\int\limits_{s^D_0}^{\infty} ds\,e^{-s/M^2}
\mbox{Im} F^{\pi,t}_{0(1)}(q^2,s).
\label{eq:subtr}
\ee
The expressions for 
$ F_{0}^{\pi,t}$ and $ F_{1}^{\pi,t}$ in (\ref{eq:LCSR})
are obtained from the corresponding 
expressions given in \cite{DKMMO} for $B\to \pi$ LCSR, 
replacing $b\to c$ and $B\to D$. 
The second LCSR for the combination $(f^+_{D\pi}+f^-_{D\pi})$ 
is obtained from the invariant amplitude 
$\widetilde{F}^\pi$ and has the same form 
as (\ref{eq:LCSR}), with the invariant amplitudes 
$\widetilde{F}_{0(1)}^{\pi,t}$, replacing 
$F_{0(1)}^{\pi,t}$, and without  
the factor $1/2$ in the coefficient.

For $D\to K$ form factors, as explained above,  
the $SU(3)_{fl}$ violation effects 
are  taken into account in both LCSR's for $f_{DK}^+$ and 
$(f^+_{DK}+f^-_{DK})$ in the LO part, keeping $p^2=m_K^2\neq 0$ 
in the hard kernels and  taking into account the $O(m_s)\sim O(m_K^2)$ effects 
in the kaon DA's. For the sake of completeness, we do not 
expand these expressions in $m_K^2 \sim m_s$, 
although it is clear that only the first-order terms of 
this expansion are important numerically.
Furthermore, having in mind that 
both $O(\alpha_s)$ and $O(m_s)$ corrections are 
reasonably small,
we do not take into account the combined 
$O(\alpha_sm_s)$ effects,
originating from nonzero $m_s$  
and $p^2=m_K^2$ in the NLO  diagrams. 
These effects demand a dedicated calculation, taking 
into account the mixing between various twist components 
at the $O(m_s)$ level. Hence, in the adopted  
approximation
$F^{K,2(3)}_{1}=F^{\pi,2(3)}_{1}$.

  In Appendix B, we present the 
expressions for all LO terms $F^{K,t=2,3,4}_{0}$ 
entering LCSR's for $D\to K$ form factors. 
In the limit $m_s\to 0$ ($m_K^2\to 0$) and $\mu_K\to \mu_\pi$  
the corresponding terms $F^{\pi,t=2,3,4}_{0}$ in the 
$D\to \pi$ LCSR's are easily reproduced. 
The expressions for the 
NLO terms of twist-2 and twist-3 are presented in \cite{DKMMO}
in a form of dispersion integrals, and we replace $m_b\to m_c$.
These very bulky formulae are not displayed here.

The LCSR's for $D\to K$ form factors   
were also compared to the recent update of the LCSR's 
for $B\to K$ form factors presented in \cite{DM}. 
The expressions presented in Appendix B agree 
with those in \cite{DM}, up to   
the twist-3 three-particle part of LCSR, which is incomplete
in \cite{DM}. We also modified 
the $m_K^2$-corrections
to the two-particle twist-4 DA's (see discussion in Appendix A),
hence, there are small differences in the corresponding 
twist-4 terms in LCSR's. Both differences have 
a minor impact on numerical results.    
Finally, we compared  our expressions with the 
LCSR for the $D\to P$ ($P=\pi,K$) 
form factor in LO at $q^2=0$ presented 
in \cite{BallDpi}, 
where only the $O(m_P^2)$ terms are retained.
In the limit $m_P^2=0$, the equations    
(\ref{eq:fplusDpiLCSRcontribTw2}), 
(\ref{eq:fplusDpiLCSRcontribTw3}), 
and (\ref{eq:fplusDpiLCSRcontribTw4}) 
are reduced to the corresponding terms  in Eq.~(2) of \cite{BallDpi}, 
except we obtain an opposite  sign of the contribution 
of twist-4 DA $\psi_{4;P}$.  We also were not able to completely
reproduce the $O(m_P^2)$ terms in this equation. 
 
The expected accuracy  of LCSR (\ref{eq:LCSR})
is determined by the uncertainties of 
the combined expansion in twists and $\alpha_s$ of the correlation function.
The choice of Borel parameter plays an important role here. 
Note that Borel transformation effectively replaces 
the powers of the denominator $D$ in the higher-twist 
contributions by the inverse powers of $M^2$, as seen from (\ref{eq:Borel}) 
\footnote{The LCSR expressions in Appendix~B are given in a compact form, 
using derivatives of DA's. To restore the inverse powers of $M^2$ 
one has to perform a partial integration.}.
Hence, although $M^2$ is an arbitrary scale,  
it should be taken sufficiently large to preserve the twist hierarchy. 
Parametrically, in the limit of the heavy $c$-quark mass, 
the relevant scaling relation is $M^2\sim 2m_c\tau$. 
Importantly, also the scale 
$\tau$ has to be much larger than the typical soft scales 
of $O(\Lambda_{QCD})$,  such as the normalization parameters 
of higher-twist DA's .  The twist expansion
is then ``protected'' by the combinations of $1/m_c$ and $1/\tau$ 
suppression factors, as one can explicitly prove by expanding 
LCSR in powers of $1/m_c$ (for more details on heavy-quark expansion
of LCSR  see \cite{BBB,KRW98,BallSCET}).
  
Another important step in the derivation of 
LCSR is the subtraction of the integral over higher states 
in (\ref{eq:subtr}).
This procedure  
introduces the lower limit $ u_0=(m_c^2-q^2)/(s_0^D-q^2)$ 
in the convolution integrals entering LCSR
(see the expressions for $F^{K,t}_{0}(q^2,M^2,s_0^D)$ in Appendix B).
In LCSR's for $D\to\pi,K$ form factors the integration limit is $u_0\simeq 0.3-0.4$.
Hence, the suppression of higher twists is not influenced,
because after subtraction, the large part of the $u$-region 
in the integrals over DA's is retained. This is not the case in the limit 
of infinitely heavy quark and small $q^2$,
when, according to the scaling rule, 
$s_0^D\sim (m_c^2+2m_c\omega)$ and $u_0\sim 1-2\omega/m_c\to 1$. In this limit 
the power suppression of the higher-twist terms in LCSR 
depends also on the end-point behavior 
of DA's, modifying the initial twist hierarchy
of OPE. For example, the formal $1/m_c$ suppression of the 
twist-3
terms mentioned above, is removed at $m_c\to \infty$. 
Still, this modification does not 
influence the numerical suppression of $t\geq 4 $ contributions, 
provided   the effective Borel scale $\tau$ is kept large.

The accuracy of the quark-hadron duality approximation
is difficult to estimate in a model-independent way. 
In LCSR, one minimizes  the sensitivity 
to this approximation, using not too large $M^2$, 
in order to suppress  the integral  
over the higher states exponentially. In addition, 
the effective threshold  $s_0^D$ is determined,   
calculating the $D$-meson mass from the differentiated LCSR
(\ref{eq:LCSR})
and adjusting the result to the measured value,
as it was done in \cite{BZ04,DKMMO}. This procedure is more
reliable than fixing the effective threshold from 
a stability of the sum rule with respect to the Borel parameter variation. On the other hand, if the result of LCSR calculation actually reveals a weak dependence on $M^2$ 
(within the adopted interval), 
that is an important indication of the reliability of the method.

\section{Choice of the input }

In this section we specify and discuss the choice of input parameters 
entering LCSR's. 
As already  mentioned, we use the $\overline{MS}$ scheme for 
the quark masses. In previous analyses, 
(e.g., in \cite{KRWWY,BallDpi}) the $c$-quark  pole mass 
was used in the final sum rule, which is certainly less 
convenient for a correlation function with 
a virtual $c$ quark.
For the $c$-quark mass value we adopt
the interval obtained from charmonium sum rules with 
$O(\alpha_s^3)$ accuracy \cite{mc}, 
\be
\bar{m}_c(\bar{m}_c)= (1.29\pm 0.03)\,\mbox{GeV},
\label{eq:mc}
\ee
where, conservatively,  we double the error.  
This interval is in a good agreement with the recent 
lattice determination \cite{Allison}.
For the $s$-quark mass we take the interval 
\be
m_s(\mu=2\,\mbox{GeV}) = (98 \pm 16)\,\mbox{MeV}\,, 
\label{eq:ms}
\ee
which  covers the range of the QCD sum rule 
determinations with $O(\alpha_s^4)$ accuracy \cite{ms}. 
Employing the well known ChPT relations \cite{Leutw}:
\be \label{eq:ChPT}
R= \frac{2m_s}{m_u+m_d}=24.4\pm 1.5,~~ 
Q^2=\frac{m_s^2-(m_u+m_d)^2/4}{m_d^2-m_u^2}=(22.7\pm 0.8)^2,
\ee  
the $u,d$-quark masses and their sum can be calculated.
Since we neglect $m_u$ and $m_d$ everywhere, except in the parameters
$\mu_\pi$ and $\mu_K$, we simply use the above
relations and obtain (adding 
the errors in quadrature):
\ba
\mu_\pi(2 ~\mbox{GeV})&=& \frac{m_\pi^2R}{2m_s(2 ~\mbox{GeV})}
=(2.43\pm 0.42)~\mbox{GeV}\,,\nonumber\\
\mu_K(2 ~\mbox{GeV}) &=& \frac{m_K^2}{m_s(2 ~\mbox{GeV})\Big[ 1+\frac{1}{R}\left(1-\frac{R^2-1}{4 Q^2}\right)
\Big]}= (2.42 \pm 0.39)~\mbox{GeV}\,,
\label{eq:mupiK}
\ea
with a remarkably small $SU(3)_{fl}$ violation.
  
In our numerical analysis,
the two-loop running for QCD coupling 
is used, with $\alpha_s(m_Z)=0.1176 \pm 0.002$ \cite{PDG}. 
The scale-dependence of the quark masses and parameters of DA's
is taken into account in one-loop approximation.
Furthermore, 
we use a uniform scale $\mu$ for all renormalizable  
parameters, with the same default value  $\mu=1.4\ \text{GeV}$, 
as in previous analyses of LCSR's; note that parametrically, 
$\mu\sim \sqrt{m_D^2-m_c^2}$.

The hadronic inputs in LCSR's include 
the hadron masses 
$m_{D^0} = 1.865\ \text{GeV}$, $m_{\pi\pm}= 139.6\ \text{MeV}$ and 
$m_{K^\pm}= 493.7\ \text{MeV}$ \cite{PDG}.   
The  pion and kaon decay 
constants, $f_\pi = 130.4\ \text{MeV}$ and $ f_K = 155.5\ \text{MeV}$ \cite{PDG}
normalize the twist-2 pion and kaon DA's, respectively.
All other parameters of twist 2,3,4 DA's
relevant for our calculation 
are collected in Table \ref{tab:parampiK} in Appendix \ref{app:inputDA}. Let us briefly comment on our choice.
Expressing the twist-2 DA's in terms of Gegenbauer 
polynomials, we adopt the intervals 
$a_2^\pi(1\,\text{GeV})=0.16\pm 0.01$, 
$a_4^\pi(1\,\text{GeV})=0.04\pm 0.01$, 
determined in \cite{DKMMO}, by fitting
the form factor calculated from LCSR to the 
measured shape of $B \to \pi$ form factor. 
These intervals are in agreement 
with the other determinations of $a_{2,4}^\pi$ 
summarized in \cite{BBL}.
For the first Gegenbauer moment of the kaon DA
we use  $a_1^K=0.10\pm 0.04 $, obtained 
in \cite{CKP} from the two-point QCD sum rules with NNLO accuracy.  Finally, 
the interval $a_2^{K}=0.25\pm 0.15$  is adopted 
\cite{KMM,BBL}.  All other Gegenbauer moments
are put to zero, the same approximation as in the previous 
LCSR analyses. In fact, as already noticed in \cite{BallDpi}, the sensitivity of the LCSR for the 
$D\to \pi$ form factor to Gegenbauer
moments is less than in the $B\to \pi$ case. 
In particular, we have checked that nonvanishing values 
of the next Gegenbauer moments $a_3^K$ and  $a_4^K$,  
at the level of  $a_1^K$ and $a_4^\pi$, respectively, 
have a small influence on the numerical results.
For the twist-3,4 pion and kaon DA's, in addition to the 
normalization parameters already specified in (\ref{eq:mupiK}) 
the set of parameters determined and updated in \cite{BBL} is used. Note that due to the smallness
of $f_{3\pi} \simeq f_{3K}$, the size of nonasymptotic 
corrections to the twist-3 DA's $\phi_{3\pi,3K}^{p,\sigma}$ is small.
Hence, it is justified to take asymptotic DA's  
in  the NLO twist-3 terms of LCSR calculated in \cite{BZ04,DKMMO}.

The remaining hadronic input in  
LCSR  (\ref{eq:LCSR}) is the decay constant $f_D$.  
In previous applications of LCSR's 
to heavy-light form factors,  e.g., in \cite{BZ04,DKMMO,KRWWY},   
the two-point QCD sum rule for the heavy-meson decay constant
was substituted in LCSR, leading to a 
partial cancellation between radiative gluon 
corrections. However, the two-point  sum rule,
with its own Borel-parameter range and effective 
duality threshold,
introduces an additional uncertainty in the calculated form factors.
Note that in the sum rules with $\overline{MS}$ heavy-quark mass
the $\alpha_s$-corrections  are not sizeable and, therefore, their
partial cancellation is not that important. On the other 
 hand, the decay constant $f_D$ has already been 
 measured in $D\to l \nu_l$ \cite{CLEOfD} with a
 very good accuracy. For that reason, in our numerical
 analysis we prefer to use the experimental result, assuming 
 the isospin symmetry: $f_{D^0}=f_{D\pm}$,  
and taking  $f_{D^+}= 205.8\pm 8.9 $\,MeV  from \cite{CLEOfD},
where we add the errors in quadrature. 
Importantly, this value is obtained  assuming $|V_{cd}|=|V_{us}|$,
with $|V_{us}|=0.2255\pm 0.0019 $ from \cite{PDG}.
Extracting $|V_{cd}|$ below, we will take this into 
account.
The two-point QCD sum rule prediction for $f_D$ 
used in previous analyses (e.g., in \cite{KRWWY}) agrees  
with the experimental interval, but has a larger uncertainty.

Finally, we specify the ``internal'' parameters 
of LCSR (\ref{eq:LCSR}): the interval of the Borel parameter  $M^2$ 
and the effective threshold $s_0^D$. For the 
former, we choose the  region 
$M^2=(4.5\pm 1.0)\ \text{GeV}^2$,
close to the one used  in \cite{KRWWY}. 
The threshold parameter $s_0^D=(7.0\pm 0.5)\ \text{GeV}^2$ 
is fixed by reproducing (within 2\% accuracy) the 
$D$-meson mass from the auxiliary sum rule, 
obtained from differentiating the LCSR 
(\ref{eq:LCSR}) over $1/M^2$ and  dividing the result
by (\ref{eq:LCSR}). 
With our choice of $M^2$ and $s_0^D$, 
the usual criteria are fulfilled for the 
LCSR (\ref{eq:LCSR}): smallness of the subleading 
twist-4 contributions ($< 5\%$ of the twist-2 term) 
and, simultaneously, 
suppression of higher state contributions 
($<10\%$ of the total correlation function). Since below 
we also calculate the form factors 
at negative $q^2$, we checked that the adopted ranges
of $M^2$ and $s_0^D$ are equally applicable for 
$-2 ~\mbox{GeV}^2\leq q^2 \leq 0$.
For the second LCSR for ($f^+_{D\pi}+f^-_{D\pi}$),  the same intervals are taken for consistency. The differences between the Borel and threshold parameters for the sum rules for $D\to\pi$ and $D\to K$  form factors turn out to be negligible. 
Note also that   
the effective Borel scale $\tau= M^2/(2m_c)\simeq
1.7\ \text{GeV}$ is sufficiently large.

\section{Form factors at $q^2=0$}\label{sec:formf_at_0}

Substituting in LCSR (\ref{eq:LCSR})
the input specified above, we calculate 
the form factors $f^+_{D\pi}(0)$ and $f^+_{DK}(0)$.
The numerical evaluation was done in two different ways:
firstly by a direct integration
over imaginary parts of hard-scattering amplitudes, 
and secondly, applying the numerically equivalent method 
of analytical continuation explained and used in \cite{DKMMO,DM}.

\begin{figure}[t]
\centering
 \includegraphics[scale=0.27]{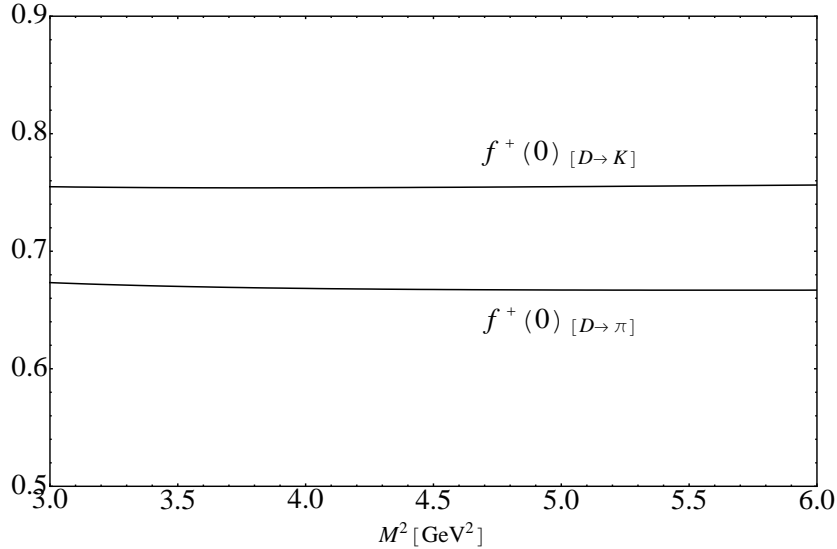}
  \caption{\it
The Borel-parameter dependence of the form factors 
$f^+_{D\pi}(0)$ and $f^+_{DK}(0)$
calculated from LCSR.
}\label{fig:Borel}
\end{figure}

\begin{figure}[t]
\centering
 \includegraphics[scale=0.27]{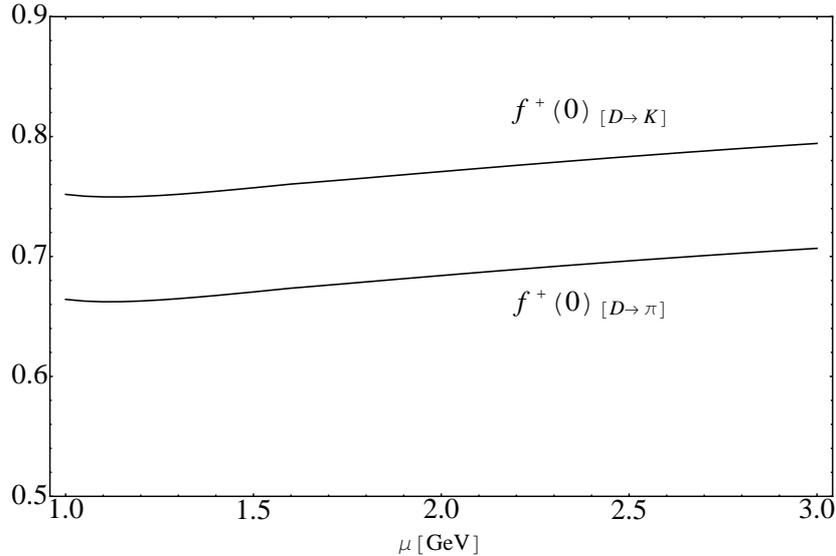}
  \caption{\it
The scale-dependence of the form factors 
$f^+_{D\pi}(0)$ and $f^+_{DK}(0)$
calculated from LCSR }\label{fig:scale}
\end{figure}

The results  at the central values of input parameters 
are displayed  in Table~\ref{tab:fplus0}. Their dependence 
on Borel parameter is shown in Fig.~\ref{fig:Borel},
and exhibits a remarkable stability, even beyond the 
adopted interval. The scale-dependence 
displayed  in Fig.~\ref{fig:scale}, is
also mild. Conservatively, we consider the variation of the 
calculated form factors 
with the  scale change in the interval $1.0<\mu<3.0$ GeV 
as one of the  uncertainties. 
\begin{table}[b]
\caption{\it Form factors $f^+_{D\pi}(0)$ and $f^+_{DK}(0)$ 
calculated from LCSR (\ref{eq:LCSR}) and the estimated uncertainties
due to the variation of the input.}
\vspace{-0.1cm}
\begin{center}
{\small
\begin{tabular}{|c|c|c|c|c|c|c|l|l|}
\hline
Formf. 
&\raisebox{-1.8ex}[0mm][0mm]{$M^2$}
&\raisebox{-1.8ex}[0mm][0mm]{$\mu$}
&\raisebox{-1.8ex}[0mm][0mm]{$s_0^D$}
&\raisebox{-1.8ex}[0mm][0mm]{$(f_D)_{exp}$}
&\raisebox{-1.8ex}[0mm][0mm]{$m_c$}
&$m_s$ 
&Gegenbauer
&tw.5\\
centr.value&&&&&&$\mu_{\pi,K}$&moments&(est.)\\[2mm]
\hline 
$f^+_{D\pi}(0)$ &&&&&&&&\\
\raisebox{-.8ex}{$0.667$}&${}^{~+0.003}_{~-0.001}$
&${}^{~+0.04}_{~-0.003}$
&$\pm 0.01$&$\pm 0.03$&${}^{~+0.005}_{~-0.006}$
&${}^{+0.08}_{-0.06}$
&$\pm 0.001$~($a^\pi_{2,4}$)
&$\pm 0.017$\\[2mm]
\hline
$f^+_{DK}(0)$&&&&&&&$\pm 0.003$ ~$(a_1^K)$ &\\
\raisebox{-.8ex}{$0.754$} &${}^{~+0.001}_{~-0.0004}$
&${}^{~+0.04}_{~-0.006}$
&$\pm 0.01$
&$\pm 0.03$&${}^{+0.005}_{-0.007}$&${}^{+0.09}_{-0.06}$&$\pm 0.03$~($a_2^K$)
&$\pm 0.001$\\
&&&&&&&$\pm 0.01$ ~$(a_{3,4}^K$) &\\
\hline
\end{tabular}
}
\end{center}
\label{tab:fplus0}
\end{table}
In addition, we investigate the numerical hierarchy
of various contributions to LCSR. The sample of results 
for $f^+_{D\pi}(q^2)$ and $f^+_{DK}(q^2)$ 
is collected in Table~\ref{tab:contrib}. The 
dominance of the twist-3 LO contribution was anticipated, due to the factor $\mu_\pi/m_c>1 $. At the same time, the 
subleading twist-4 contributions are numerically strongly suppressed. The 
NLO corrections to twist-2,3 terms are also small, 
a clear indication that the "soft-overlap" mechanism 
dominates in $D\to \pi,K$ form factors.

Furthermore, we estimate  separate uncertainties  of our 
calculation by  varying each input parameter within 
its allowed interval.  
All significant uncertainties of $f^+_{D\pi(K)}(0)$ are collected in Table~\ref{tab:fplus0}. (For brevity we do not show a similar table for $f^+_{D\pi(K)}(0)+f^-_{D\pi(K)}(0)$, presenting only the total uncertainty below.) The remaining small
effects, e.g., due to the variation of 
$\alpha_s$, are not shown, but included in the total uncertainty.
Note that, according to (\ref{eq:mupiK}),
the error related to both $\mu_{\pi}$ 
and $\mu_K$ is influenced by 
the uncertainties in the determination of $m_s$ and ChPT parameters.
To roughly estimate the effects of the unknown twist-5 
contributions to LCSR's, we assume  that their ratio to the 
twist-3 contribution  is equal to the calculated ratio 
of twist-4  and twist-2 (LO) terms. Since in the sum rules for $f_{D\pi(K)}^+ + f_{D\pi(K)}^-$ there is no LO twist-2 contribution, we 
conservatively assume that the magnitudes of the twist-5 and twist-4 terms 
are the same.  In addition, 
to assess the effect of higher Gegenbauer moments $a_{3}^K$ 
and $a_4^K$ on the $D\to K$ form factor, we 
recalculated this form factor assuming 
$a_{3}^K= \pm a_{1}^K$ and $a_{4}^K= \pm a_{4}^\pi$.
The small variations due to the abovementioned 
effects are treated as separate uncertainties and included
in the error budget in Table~\ref{tab:fplus0}.
\begin{table}
\caption{\it Relative contributions  
to LCSR for the form factors $f^+_{D\pi}(q^2)$ and $f^+_{DK}(q^2)$ 
}
\vspace{-0.1cm}
\begin{center}
\begin{tabular}{|c|c|c|c|c|}
\hline
Contribution &$f^+_{D\pi}(0)$&$f^+_{D\pi}(q^2=-2~\mbox{GeV}^2)$
&$f^+_{DK}(0)$&$f^+_{DK}(q^2=-2 ~\mbox{GeV}^2)$ \\\hline
tw2 LO & 35.9\%& 32.9\% & 36.2\% & 33.7\% \\
tw2 NLO &  6.3\%& 8.4\% & 6.0\% & 7.9\% \\
\hline
tw3 LO & 66.0\%& 59.5\% & 67.9\% & 58.9\% \\
tw3 NLO &-9.5\%& -2.9\% & -10.1\% & -2.9\% \\
\hline
tw4 LO &1.4\%& 2.2\% & -0.07\% & 2.4\% \\
\hline
\end{tabular}
\end{center}
\label{tab:contrib}
\end{table}

Adding all uncertainties in quadrature, we obtain:
\ba
f^+_{D\pi}(0)= 0.67^{+0.10}_{-0.07},
\label{eq:fDpi0}\\  
f^+_{DK}(0)=0.75^{+0.11}_{-0.08}\,.
\label{eq:fDK0} 
\ea

The second LCSR yields numerical results for 
the sums of the form factors:
\ba
f^+_{D\pi}(0) + f^-_{D\pi}(0)= 0.46^{+0.12}_{-0.09},
\label{eq:fDpiplmin0}\\  
f^+_{DK}(0)+f^-_{DK}(0)=0.60^{+0.12}_{-0.09},
\label{eq:fDKplmin0} 
\ea
We also quote our prediction for the products
of $D$ decay constant and the form factors: 
\ba
f_Df^+_{D\pi}(0)= 137^{+19}_{-14} ~\mbox{MeV},
\label{eq:prodpi}
\\  
f_Df^+_{DK}(0)=155^{+21}_{-15}~\mbox{MeV}.
\label{eq:prodK}
\ea
These quantities are independent of the experimental 
value of $f_D$, 
used to calculate (\ref{eq:fDpi0}) and  (\ref{eq:fDK0}), 
and therefore have a slightly smaller uncertainty.

Finally, the predicted ratio of the form factors  is 
\be
\frac{f^+_{D\pi}(0)}{f^+_{DK}(0)}=
0.88 \pm 0.05\,,
\label{eq:ratio}
\ee
where the $f_D$ dependence drops out and 
some uncertainties largely cancel. 

In Table~\ref{tab:compar} we compare theoretical
predictions for the form factors 
$f^+_{D\pi}(0)$ and $f^+_{DK}(0)$,
obtained in lattice  QCD and from LCSR's.
Our results are in a good agreement with the
lattice determinations. 
The form factors  (\ref{eq:fDpi0}) and  (\ref{eq:fDK0}) 
and their ratio  (\ref{eq:ratio}) are also in accordance  with the previous LCSR estimates
\cite{KRWWY,BallDpi}.
The $D\to K$ form factor 
is now more accurately determined  than in  
\cite{KRWWY}, due to a better knowledge of the 
$c$- and $s$-quark masses and of the parameter $a_1^K$. 

\begin{table}[b]
\caption{\it Comparison of theoretical predictions 
for the form factors $f^+_{D\pi}(0)$ and $f^+_{DK}(0).$ 
}
\vspace{-0.1cm}
\begin{center}
\begin{tabular}{|cc|c|c|}
\hline
Method &[Ref.] &$f^+_{D\pi}(0)$&$f^+_{DK}(0)$\\
\hline
Lattice QCD &\cite{APE01}&$0.57\pm 0.06\pm 0.02$& $0.66\pm 0.04\pm 0.01$ \\
&\cite{Aubin05}&$0.64\pm 0.03\pm 0.06$&$0.73\pm 0.03\pm 0.07$\\
&\cite{QCDSF09}&$0.74\pm 0.06\pm 0.04$&$0.78\pm 0.05\pm 0.04$\\
\hline
LCSR&\cite{KRWWY}&$0.65\pm 0.11$&$0.78^{+0.2}_{-0.15}$\\
&\cite{BallDpi}&$0.63\pm 0.11$&$0.75\pm 0.12$\\
&this work &$0.67^{+0.10}_{-0.07}$&$0.75^{+0.11}_{-0.08}$\\
\hline
\end{tabular}
\end{center}
\label{tab:compar}
\end{table}

In the final part of this section we 
present our predictions for the slopes and ratios 
of the form factors at $q^2=0$, that is, at large recoil of the pion
or kaon,  (see \cite{Hill} for definition). 
We start with the parameter $\delta$ 
for $D\to \pi$ transitions,
which is simply calculated by taking the ratio of two LCSR's:
\be
\delta_{D\pi}=1+\frac{f^{-}_{D\pi}(0)}{f^+_{D\pi}(0)}= 
0.69 \pm 0.09\,.
\label{eq:delta}
\ee
The slope of the scalar form factor  at $q^2=0$
normalized by $f^+_{D\pi}(0)$  
is another interesting characteristics of the form factor. 
We obtain: 
\be
\beta_{D\pi}=\left[\left(\frac{m_D^2-m_\pi^2}{f^+_{D\pi}(0)}\right)
\frac{df^0_{D\pi}(q^2)}{dq^2}\Big|_{q^2=0}\right]^{-1}= 1.4 \pm 0.3\,. 
\label{eq:beta}
\ee
Combining these two parameters, we are able to predict
the combination
\be
1+1/\beta_{D\pi} -\delta_{D\pi}=1.02 \pm 0.18, 
\label{eq:combDpi}
\ee
consistent with the CLEO measurement \cite{CLEO2}: 
$1+1/\beta_{D\pi} -\delta_{D\pi}=
0.93\pm 0.09 \pm 0.01$. 
The corresponding slope parameters for $D\to K$ form factors 
predicted from LCSR are:
\be
\delta_{DK}= 0.79 \pm 0.07,~~ \beta_{DK}= 1.6 \pm 0.4\,,
\ee
so that
\be
1+1/\beta_{DK}-\delta_{DK}=0.84 \pm 0.17\,
\label{eq:combDK}
\ee
also agrees with 
the experimental result \cite{CLEO2}: 
$1+1/\beta_{DK}-\delta_{DK}=
0.89\pm 0.04\pm0.01$. 

As discussed in the literature
(see e.g, \cite{Hill}), the values of 
$\delta$ and $\beta$ for heavy-light form factors 
reflect the proportion of their hard-scattering 
and soft-recoil components  and, respectively, 
their deviation from the scaling behavior predicted in the 
combined heavy-quark and large-recoil limit.
We postpone a more detailed discussion of these parameters 
to a future study.

\section{Determination  of $|V_{cd}|$ and $|V_{cs}|$}

The latest CLEO measurements of semileptonic charm decays 
\cite{CLEO2},  fitted to  various form factor parameterizations 
(to be discussed in detail in the next section), yield   the products 
$f_{D\pi}(0)|V_{cd}|$ and $f_{DK}(0)|V_{cs}|$.
Both CKM matrix elements can now be determined using our predictions for the form factors  $f_{D\pi}(0)$  and  $f_{DK}(0)$.

Having an accurate experimental 
value for $f_D$ at our disposal, allows us to make the extraction 
 of $|V_{cd}|$ less dependent on the theoretical uncertainty
 of LCSR, than in previous analyses, where a 
 sum rule prediction for $f_D$ was used. We employ the 
 CLEO result \cite{CLEOfD} for the  $D$-meson decay 
 constant  multiplied by $|V_{cd}|$, (i.e,
without  the additional assumption 
$|V_{cd}|=|V_{us}|$): 
\be
f_D|V_{cd}|=46.4 \pm 2.0 \mbox{ MeV}.
\label{eq:fDVcdexp}
\ee
On the other hand,  the CLEO result \cite{CLEO2} 
for the same CKM  
matrix element multiplied by  the form factor is : 
\be
f_{D\pi}(0)|V_{cd}|= 0.150\pm 0.004\pm 0.001\,,
\label{eq:fDpiVcdexp}
\ee
obtained from the fit of the  $q^2$-bins in 
$D\to \pi e\nu$ in a form of the series parameterization. 
The product of the above two 
experimental numbers is then divided by the LCSR prediction
(\ref{eq:prodpi}), yielding $ |V_{cd}|^2$, from which  we obtain:   
\be
|V_{cd}|
=0.225\pm 0.005 \pm 0.003\ ^{+0.016}_{-0.012}\,,  
\label{eq:vcd}
\ee
where the first and second errors 
originate from 
the experimental errors in (\ref{eq:fDVcdexp}) and 
(\ref{eq:fDpiVcdexp}), respectively, and the third error
is due to the uncertainty of LCSR.
Note that this procedure involves the square of 
$|V_{cd}|$ on the experimental side. Hence, the 
theoretical uncertainty of LCSR given in (\ref{eq:prodpi}) approximately 
halves in (\ref{eq:vcd}). Our result is in a good agreement 
with the value 
$|V_{cd}|=0.234\pm 0.007\pm 0.002\pm 0.025$
determined in \cite{CLEO2} by
using the lattice QCD value of $f^+_{D\pi}(0)$ from \cite{Aubin05}. 
This agreement  is not surprising because the form factor 
obtained from LCSR is close to the lattice result 
(see Table~\ref{tab:compar}).

Furthermore, we determine the ratio of $|V_{cd}|$
to  $|V_{cs}|$, dividing (\ref{eq:fDpiVcdexp}) 
by 
\be
f_{DK}(0)|V_{cs}|= 0.719\pm 0.006\pm 0.005,
\label{eq:fDKVcsexp}
\ee
obtained from $D\to K e \nu_e$ data fit \cite{CLEO2}.
Using the ratio (\ref{eq:ratio}) we obtain:
\be
\frac{|V_{cd}|}{|V_{cs}|}= 0.236\pm 0.006\pm 0.003\pm0.013\,,
\label{eq:vcsvcd}
\ee
where the first and second uncertainties are due to the combined
(in quadratures) errors in (\ref{eq:fDpiVcdexp}) and 
(\ref{eq:fDKVcsexp}), respectively, and the third
uncertainty stems from the LCSR calculation.
Within errors,  this ratio is in agreement  
with  $|V_{cd}|/|V_{cs}|$ obtained from the 
values quoted in \cite{CLEO2}, where   
$|V_{cs}|=0.985 \pm0.009\pm0.006\pm 0.103$
was determined  using the form factor $f^+_{DK}(0)$ from 
lattice QCD \cite{Aubin05}.
Our determinations (\ref{eq:vcd}) and (\ref{eq:vcsvcd}) are consistent 
with $|V_{cd}|=|V_{us}|$ and $|V_{cs}|= |V_{ud}|$.

\section{Form factors and their shapes at $q^2\neq 0$}

The form factors $f^+_{D\pi}(q^2)$  and $f^+_{DK}(q^2)$  
are analytic functions of 
the complex variable  $q^2$. The singularities 
located on the real positive axis include the poles 
of the ground-state vector mesons $D^*$ and $D_s^*$, 
respectively,
and their radially excited states. In addition,  there  are
branch points, generated by the 
hadronic continuum states, starting from $D\pi$ and $DK$ 
thresholds at 
$q^2=(m_D+m_\pi)^2$ and $q^2=(m_D+m_K)^2$, respectively.
(Note that the $D_s\pi$ intermediate state in the $D\to K$ 
form factor is forbidden
in the isospin limit). To obtain convenient
parameterizations for the form factors, one employs 
analyticity in two different ways.

The first approach uses the dispersion relation, e.g., for the $D\to \pi$ 
form factor: 
\be
f^+_{D\pi}(q^2)= \frac{c_{D^*}}{1-q^2/m^2_{D^*}}+
\int\limits_{(m_D+m_\pi)^2}^\infty\!\!\!ds\, \frac{\rho^{D\pi}_h(s)}{s-q^2}\,,
\label{eq:disp}
\ee
where the ground-state $D^*$-pole 
at $m_{D^*}^2= (2.01\,\mbox{GeV})^2$ \cite{PDG} is isolated, 
and the hadronic spectral density $\rho^{D\pi}_h(s)$ 
includes all other intermediate hadronic states
with the $D^*$ quantum numbers.
In fact,  $m^2_{D^*}$ is slightly larger than the 
continuum threshold (so that the $D^*\to D\pi$ decay is observable), whereas in the  
dispersion relation for $f^+_{DK}(q^2)$ the $D^*_s$ pole 
at $m_{D_s^*}^2= (2.112\,\mbox{GeV})^2$ \cite{PDG}  lies below the threshold. 
Importantly, the dispersion relation (\ref{eq:disp}) has no 
subtractions, due to the expected QCD asymptotics of 
the form factor $\lim_{|q^2|\to \infty} f^+_{D\pi}(q^2)\sim 1/|q^2|$.     
The residue of the pole in (\ref{eq:disp}) is normalized as: 
\be
 c_{D^*}= \frac{f_{D^*}g_{D^*D\pi}}{2m_{D^*}}  \,,
\ee
where $f_{D^*}$ and $g_{D^*D\pi}$ are the $D^*$ decay constant
and $D^*D\pi$ strong coupling, respectively, defined in the standard way  (see e.g., \cite{BBKR}).

The rigorous dispersion relation (\ref{eq:disp}) is valid at any $q^2$. 
Hence, matching  a calculated  
form factor, e.g., the one obtained  from LCSR,
to the dispersion relation
in the region where this calculation is valid, 
one can, in principle, predict the form factor 
outside this region.  
This is, however, only possible, if the  complicated 
integral over the hadronic spectral density 
in (\ref{eq:disp}) is parameterized in a simple 
and reliable way. One possibility is to replace this integral 
by an effective pole:
\be
f^+_{D\pi}(q^2)= c_{D^*}\Big(\frac{1}{1-q^2/m_{D^*}^2}
-\frac{\alpha_{D\pi}}{1-q^2/(\gamma_{D\pi}m_{D^*}^2)}\Big)\,,
\label{eq:2pole}
\ee
where  $\alpha_{D\pi}$ and $ \gamma_{D\pi}$
parameterize the residue and the position of this pole, so that 
the normalization of the form factor at $q^2=0$ is:
\be
f^+_{D\pi}(0)=c_{D^*}(1-\alpha_{D\pi})\,.
\label{eq:2pole_normalization}
\ee

Using the relation $\gamma_{D\pi}=1/\alpha_{D\pi} $
inspired by the combined heavy-quark and large-recoil
limit, the two-pole ansatz is reduced \cite{BK} to the 
specific BK-parameterization: 
\be
f^+_{D\pi}(q^2)=\frac{f^+_{D\pi}(0)}{(1-q^2/m^2_{D^*})
(1-\alpha_{D\pi}q^2/m^2_{D^*})}\,.
\label{eq:BK}
\ee

The second approach based on the analyticity of the form factors
employs the conformal mapping of the $q^2$-plane
(see e.g., \cite{confmap} for the early uses):
\begin{equation}
 z(q^2,t_0)=\frac{\sqrt{t_+-q^2}-\sqrt{t_+-t_0}}{\sqrt{t_+-q^2}+\sqrt{t_+-t_0}},
 \label{eq:z}
\end{equation}
where $t_{\pm}=(m_D\pm m_{\pi(K)})^2$ and  
$t_0< t_+$ is an auxiliary parameter. Applying this transformation,  
one maps the $q^2$-plane (with a cut  along the positive axis) onto 
the inner part of the unit circle 
in the $z$ plane, so that 
$f_{D\pi(K)}(q^2)_{q^2\to z}$ is free from singularities   at $|z|<1$.  
The conformal mapping (\ref{eq:z}) was employed while deriving 
the unitarity bounds for the heavy-light form factors in 
\cite{zparam,BecherHill}. 
Independent of these bounds, with an optimal choice of $t_0$, the 
semileptonic region  $0<q^2<(m_D-m_{\pi(K)})^2$ is mapped onto 
the interval of small $|z|$. Hence, a simple expansion 
in powers of $z$ around $z=0$, retaining  only a few first terms, 
provides a reasonably accurate 
parameterization  of the form factor.

In this paper we will use
the recently suggested version \cite{BCL} of the  
series parameterization for $B\to \pi$ form factor. 
Adapting it to the case of $D\to \pi$ transition, we have: 
\begin{equation}
 f^+_{D\pi}(q^2) = \frac{1}{1-q^2/m_{D^*}^2}\sum\limits_{k=0}^{N} 
\widetilde{b}_k\,[z(q^2,t_0)]^k\,.
\end{equation}
As explained in \cite{BCL}, this parameterization  
ensures general analytic properties of 
the form factor: the $D^*$-pole, branch point at 
$q^2=t_{+}$ and $\sim 1/q^2$ asymptotics at large $q^2$. Furthermore,
to obey the expected near-threshold behaviour, 
the relation 
\be
\widetilde{b}_N = 
-\frac{(-1)^N}{N}\sum\limits^{N-1}_{k=0}(-1)^k\,k\,
\widetilde{b}_k
\label{eq:BCLrel}
\ee
 has to be 
introduced, reducing the number of independent parameters 
by one. In addition, we find it more convenient 
to keep the form factor at zero momentum transfer $f^+_{D\pi}(0)$  
as one of the independent parameters, correspondingly 
rescaling the coefficients in the series expansion:
$ \widetilde{b}_k=f^+_{D\pi}(0)b_k$, so that 
\be
b_0 = 1 - \sum\limits_{k=1}^{N-1} b_k\,\left[z(0,t_0)^k-(-1)^{k-N}\frac{k}{N}z(0,t_0)^N\right]\,.
\ee
This leads to the final form of the series parameterization
used in our analysis:
\ba
f^+_{D\pi}(q^2) = \frac{f^+_{D\pi}(0)}{1-q^2/m_{D^*}^2}
\Bigg\{1+\sum\limits_{k=1}^{N-1}b_k\,\Bigg(z(q^2,t_0)^k-
z(0,t_0)^k
\nonumber\\
-(-1)^{k-N}\frac{k}{N}\bigg[z(q^2,t_0)^N-
z(0,t_0)^N\bigg]\Bigg)\Bigg\}
\label{eq:BCLfp0}\,.
\ea
In \cite{BCL} the advantages of this choice with respect 
to the previous versions \cite{zparam} are discussed
(see also \cite{DGLY}).  
The analogous  ansatz for $f^+_{DK}$ contains $m_{D_s^*}^2$ in the pole 
prefactor. Note that the formal prescription for the 
conformal mapping is to multiply the r.h.s. of (\ref{eq:BCLfp0}) 
by the pole factor 
if the ground-state pole lies below the threshold $t_{+}$, 
which is, strictly speaking, only valid  for $D\to K$ form factor.  
We prefer to retain this factor in (\ref{eq:BCLfp0}) also 
for $D\to \pi$  case, 
having in mind that $D^*$ is located very close to 
the $D\pi$ threshold. In what follows, we
simply rely on the smallness of the variable $z$ 
in (\ref{eq:BCLfp0}), 
not taking into account the unitarity bounds \cite{zparam} 
for the coefficients $b_k$, because, at least for small $N$, 
these bounds are not restrictive \cite{BCL} 
(see also \cite{BecherHill}).

In order to match the  LCSR prediction for $D\to\pi$ and $D\to K$  
form factors to one of the  parameterizations
discussed above, we have to calculate  these form factors   
beyond $q^2= 0$. However, the small part 
$0\leq q^2\ll m_c^2 $  of the semileptonic   
region where our calculation is valid, 
is too narrow to serve as a "lever arm"  for fitting various parameterizations. 

In this work, in order to enlarge 
the interval of $q^2$, 
we calculate the form factors at $q^2_{min}<q^2<0$, that is, at 
negative momentum transfers not accessible
in semileptonic decays \footnote{In fact, this region
corresponds to a hypothetical (but still physical) 
process of $l \pi (K)\to \nu_l D $ scattering.}. Fitting the 
LCSR predictions in this region   
to a certain parameterization, we then 
use the analyticity of the form factors, continuing the 
fitted parameterization
to positive $q^2$ and accessing the whole semileptonic
region $0\leq q^2<t_{-}$.   

Note  that LCSR's are fully applicable at $q^2<0$, 
since the virtuality of the $c$ quark in the correlation function 
is even larger at  $q^2<0$, than at $q^2=0$. For our purpose it is sufficient 
to take  $q^2$ not too large,  and the actual numerical calculation 
is done up to  $q^2_{min}= -2 ~\mbox{GeV}^2$. In fact, there 
are several 
reasons  to keep moderate values of $|q^2|$. 
First, we still can use the same ranges of the sum rule
parameters $M^2$, $s_0^D$, $\mu$, as  specified in Sect. 3,
whereas at very large $|q^2|$ some of the choices have to be 
modified.  In addition,  at large virtualities,
$|q^2| \gg \mu^2,m_c^2$,  large logarithms in NLO terms can  
destroy the balance of perturbative expansion.
Finally, at very large negative $q^2$,  the lower limit 
of integration $u_0$ in LCSR moves too close to 1, and 
this may potentially influence the twist expansion. 

Turning to the fit procedure, we fix
the $q^2=0$  values of the form factors  
(\ref{eq:fDpi0}) and (\ref{eq:fDK0}) 
obtained from LCSR's  and, in addition, calculate 
the shapes of the form factors $f_{D\pi(K)}^+(q^2)/f_{D\pi(K)}^+(0)$
at $-2 ~\mbox{GeV}^2<q^2\leq 0$.
The uncertainty is determined in 
the same way as described in Sect.~4 for the 
form factors at $q^2=0$. The calculated $D\to \pi$ and $D\to K$ 
form factors and their shapes at $q^2\leq 0$  
are displayed in Figs. 3 and 4,
respectively. Note that the shapes have smaller uncertainties
than the form factors themselves, because in the ratios some
uncertainties cancel (e.g., the one due to $f_D$). 
\begin{figure}[t!]
\centering
 \includegraphics[scale=0.27]{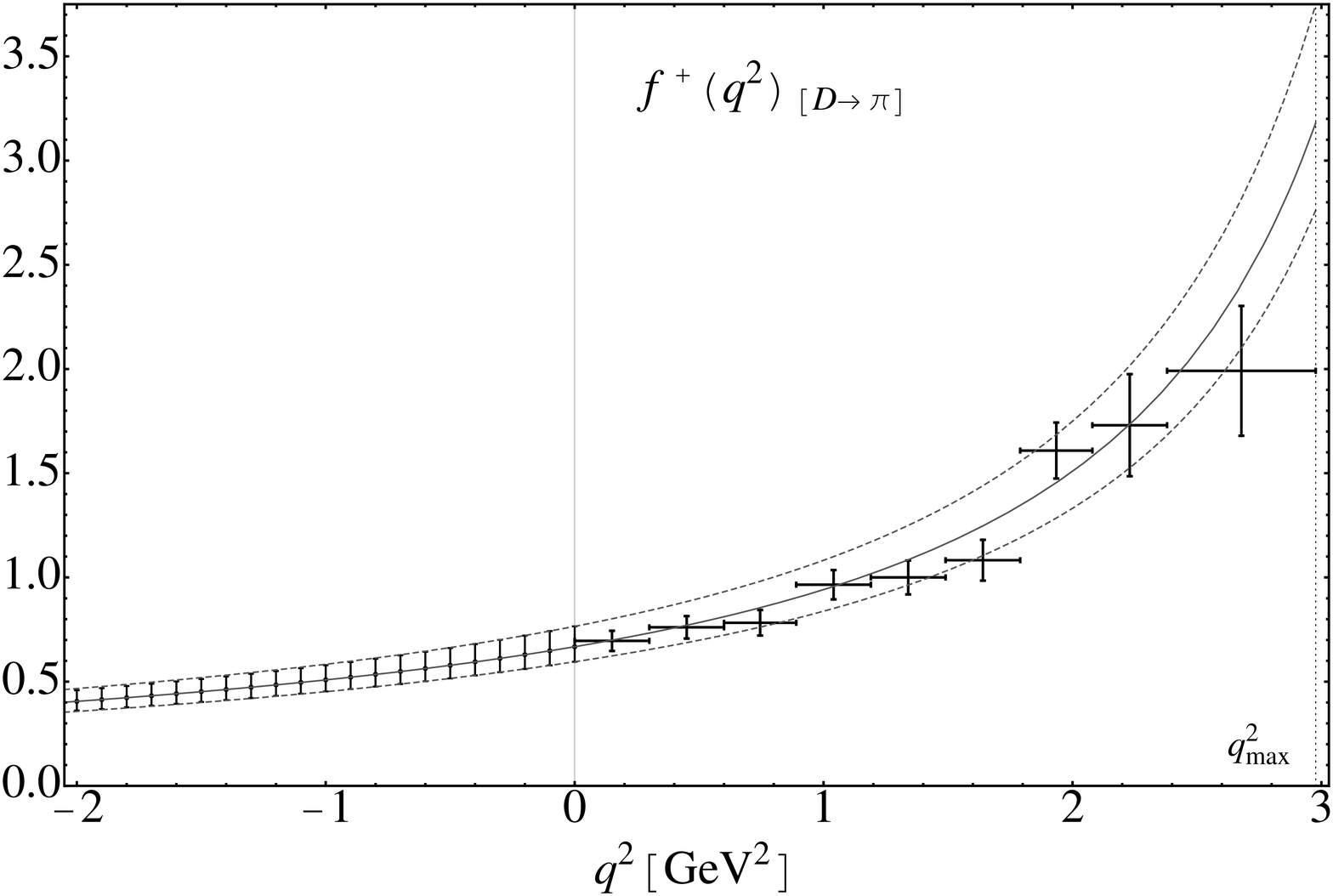}\\[3mm] 
\includegraphics[scale=0.27]{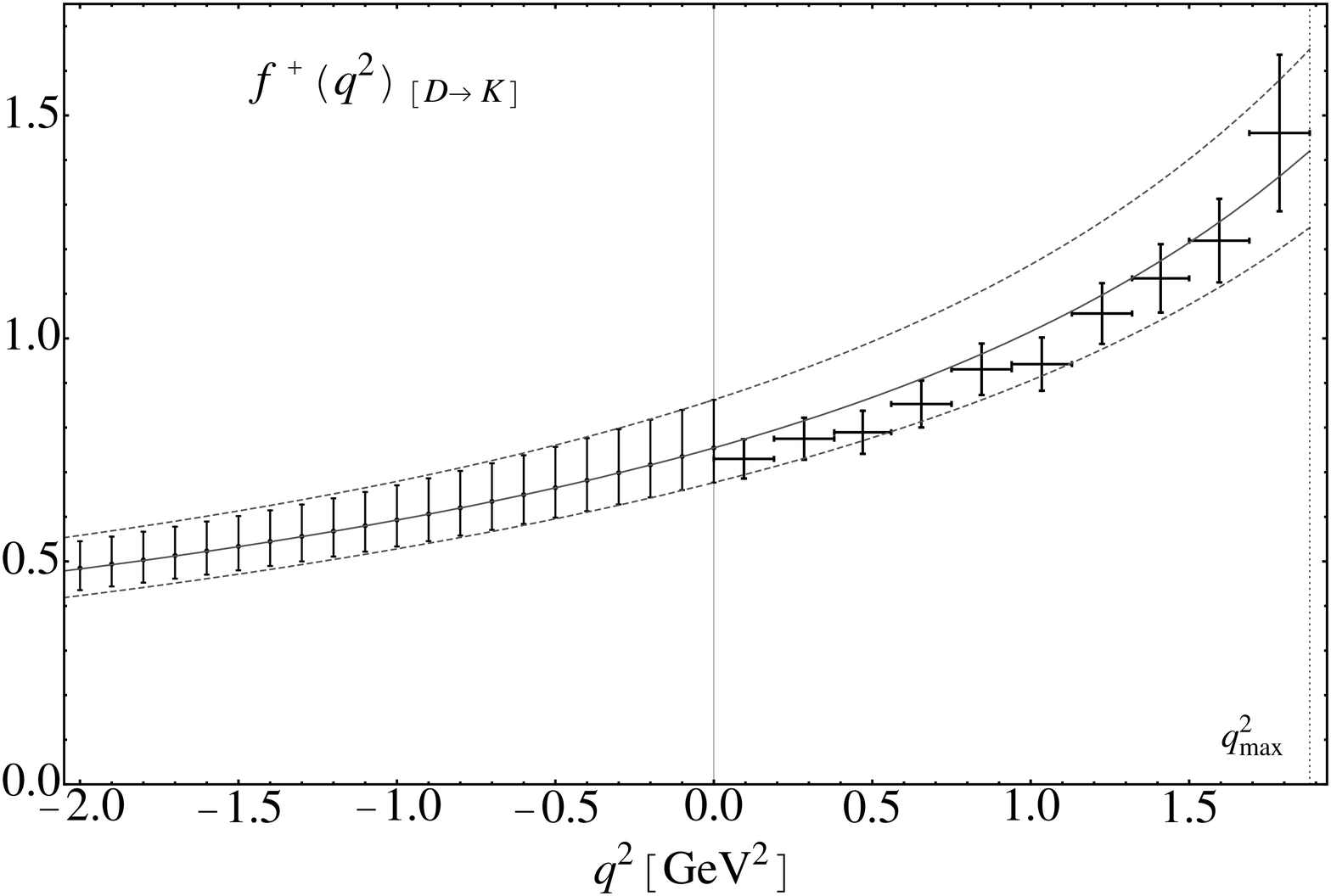}
\caption{\it Form factors $f^+_{D\pi}(q^2)$ (upper panel) 
and $f^+_{DK}(q^2)$ (lower panel).
The LCSR results with uncertainties (points with error bars at $q^2<0$) 
are fitted to series-parameterization 
(solid line and dashed lines indicating uncertainties)
and compared to the CLEO measurements \cite{CLEO1}
(points with error bars at $q^2>0$), where $|V_{cd}|$ and $|V_{cs}|$ 
are taken from \cite{PDG}.}\label{fig:ffvsCLEO}
\end{figure}

The calculated shapes, taking into account their 
uncertainties, are then fitted to  
the BK and series parameterizations, presented in 
(\ref{eq:BK}) and (\ref{eq:BCLfp0}), respectively. 
In the series  parameterization (\ref{eq:BCLfp0})
we choose $t_0 = t_+-\sqrt{t_+-t_-}\sqrt{t_+-q^2_{min}}$, 
so that  the whole interval $q^2_{min}=-2\ \text{GeV}^2<q^2\leq t_-$  
relevant for the calculation and subsequent continuation of the 
form factor is mapped onto the narrowest possible 
interval  $|z|<0.22$ ($|z|<0.09$) for $D\to\pi$ ($D\to K$).
We found that both parameterizations describe 
the shapes  calculated at $q^2\leq0$ reasonably well. 
In addition, we also fitted 
the form factors to the simplest one-pole parameterization, 
but the fits yield an unnaturally small, lower than $m_{D^*_{(s)}}$, 
pole mass. Moreover, after continuing to positive $q^2$,
the one-pole form factor noticeably deviates from
the experimentally measured shape.

The parameters of BK parameterization  
$\alpha_{D\pi}$ and $\alpha_{DK}$ 
obtained from our fit are shown in Table~\ref{tab:fitresults},
in comparison with the experimental \cite{CLEO2} 
and lattice QCD \cite{Aubin05} BK-fits.  
\begin{table}[h]
\caption{\it The shape parameters of BK parametrization}
\vspace{-0.1cm}
\begin{center}
\begin{tabular}{|c|c|c|c|}
\hline
Method & Ref & $f^+_{D\pi}(q^2)$&$f^+_{DK}(q^2)$\\
\hline
LCSR at $q^2\leq 0$  &this work &$\alpha_{D\pi} = 0.21^{+0.11}_{-0.07}$ & $\alpha_{DK} = 0.17^{+0.16}_{-0.13}$\\
\hline
Experiment & \cite{CLEO2}&
$\alpha_{D\pi}=0.21\pm 0.07\pm 0.02$ & 
$\alpha_{DK}=0.30\pm 0.03\pm 0.01$\\
\hline
Lattice QCD &\cite{Aubin05} &      
$\alpha_{D\pi}=0.44\pm 0.04\pm 0.07$&
$\alpha_{DK}=0.50\pm 0.04\pm 0.07$\\
\hline
\end{tabular}
\end{center}
\label{tab:fitresults}
\end{table}
Previous LCSR estimates of these parameters \cite{KRWWY} 
are smaller but have also larger errors; in fact, they have been 
determined from a different procedure, where, in addition to 
the LCSR form factor  at $q^2=0$, 
the calculated $D^*D\pi$ coupling was used, adding its own 
uncertainty. 

For the series parameterization (\ref{eq:BCLfp0}), 
a fit is possible already at $N=2$, i.e., with only 
one free parameter $b_1$ for the shape ($b_2$ is fixed 
from the condition (\ref{eq:BCLrel})). We obtain: 
\be
b^{D\pi}_1 = -0.8^{+0.3}_{-0.4}\,,~~ 
b^{DK}_1 = -0.9^{+0.7}_{-0.8}\,.
\label{eq:zfitfpl} 
\ee
Fits at $N=3,4$ were also performed, yielding 
numerically very close results. For $N\geq 5$ the unitarity bounds 
\cite{BCL} start to constrain the coefficients $b_k$. 

\begin{figure}[t!]
\centering
 \includegraphics[scale=0.27]{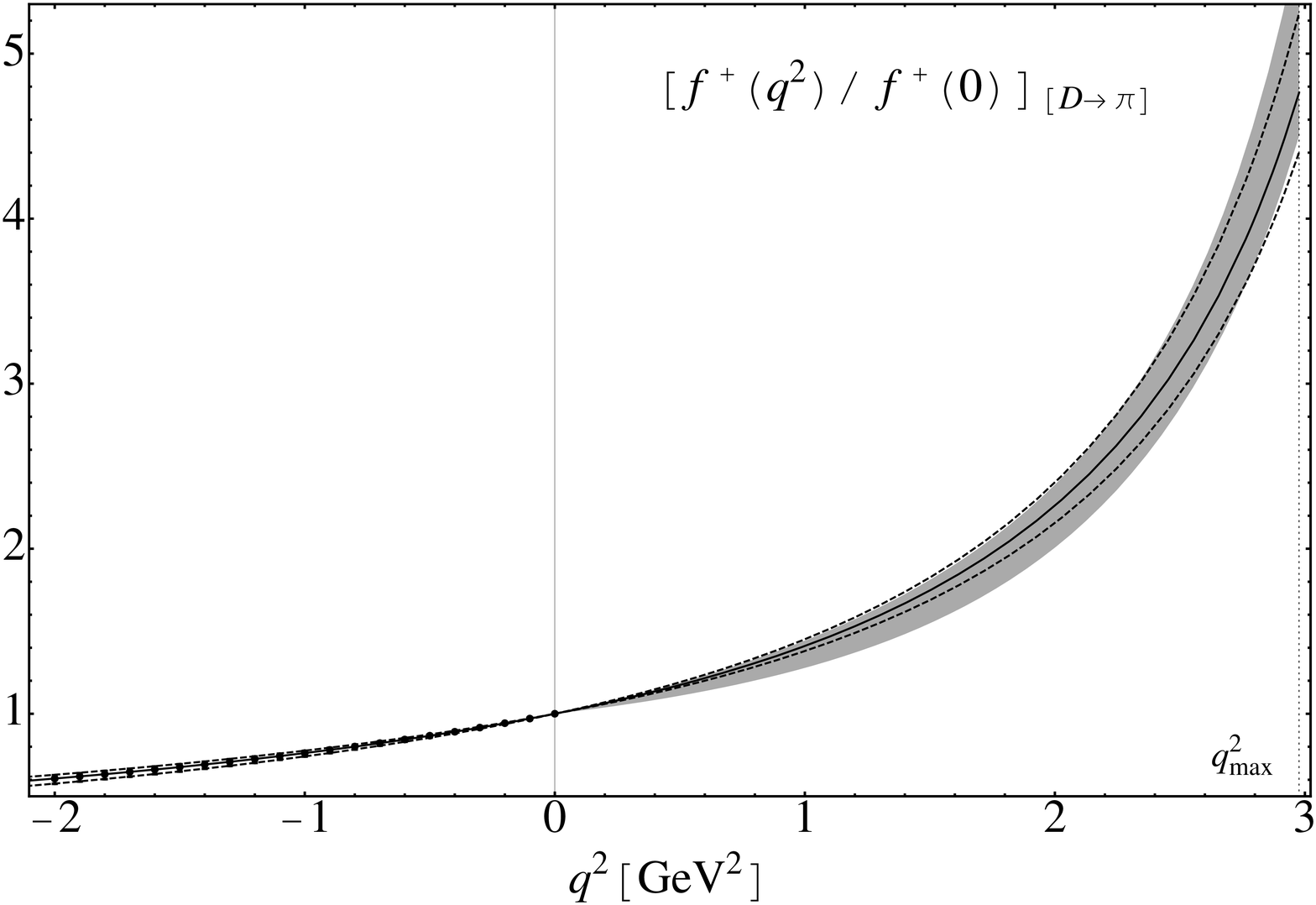}\\[3mm]
 \includegraphics[scale=0.27]{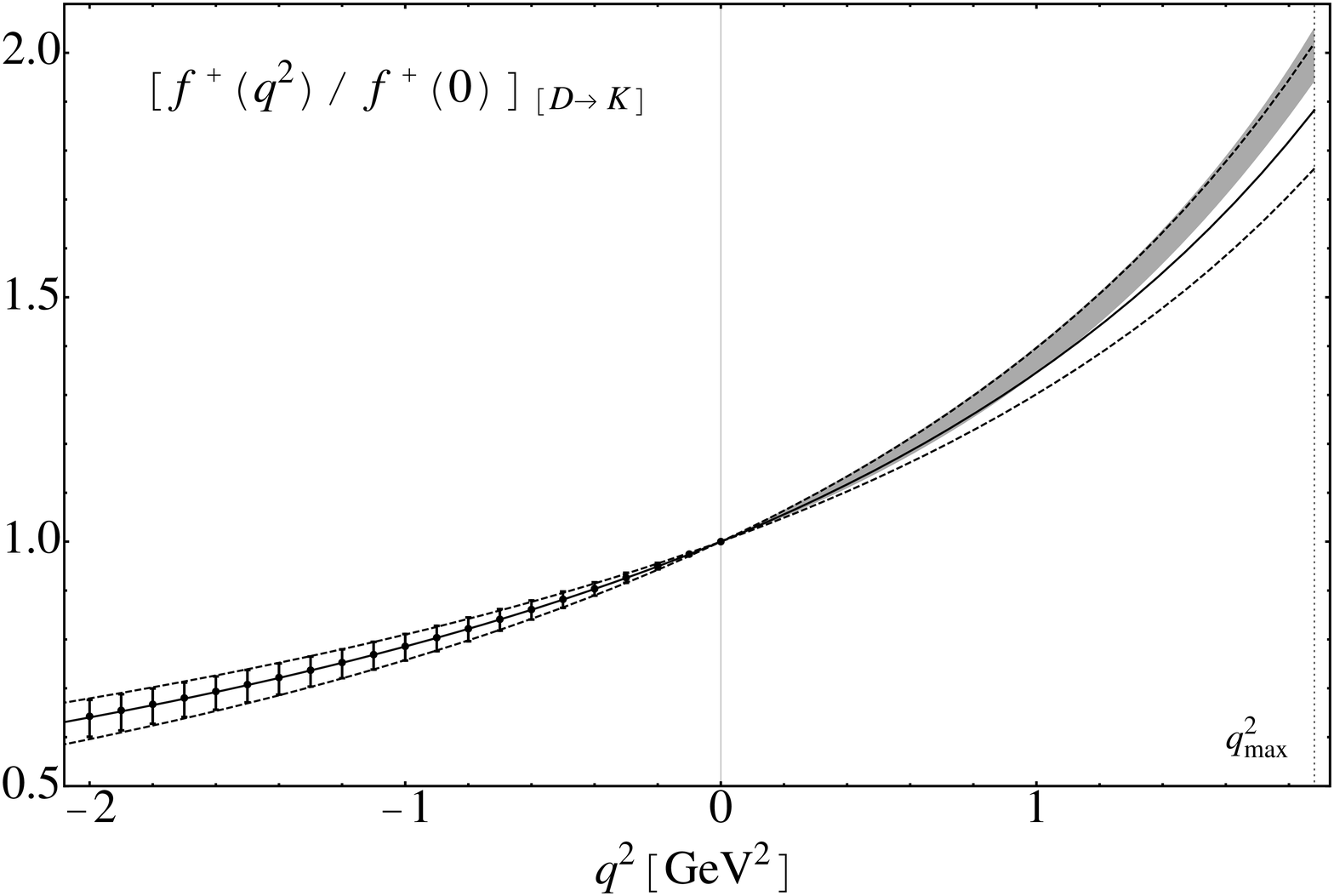}
 \caption{\it
The shapes of the $D\to \pi$ (upper panel) and $D\to K$ 
(lower panel) form factors obtained from LCSR's  
at negative $q^2$ (points with error bars), fitted 
to the series parameterization (solid line and dashed lines 
indicating uncertainties) and 
compared to the shapes measured by CLEO \cite{CLEO2}
(shaded regions).}
\label{fig:ffslopevsCLEO}
\end{figure}

In what follows, we choose the $N=2$
series parameterization to be our preferred 
analytic expression for the shape, having in mind that 
it is less model-dependent than the effective pole
ansatz for the dispersion integral.  
Continuing this parameterization to the semileptonic region 
$q^2\leq t_-=(m_D-m_{\pi(K)})^2= 2.98 ~\mbox{GeV}^2(1.88 
~\mbox{GeV}^2)$,  
we compare in Fig.~\ref{fig:ffvsCLEO} the form factors with the 
experimentally measured ones, presented in \cite{CLEO1} 
in $q^2$-bins. For the normalization 
of the  data we take the averages: 
$|V_{cd}|=0.230\pm 0.011$ and $|V_{cs}|=1.04\pm 0.06 $ from \cite{PDG}. 
The form factor shapes, which are independent of 
normalization at $q^2=0$ and CKM parameters, 
are displayed in Fig.~\ref{fig:ffslopevsCLEO}. 
We compare  our predictions for the series parameterization  
with the shapes  obtained in \cite{CLEO2} 
and observe a good agreement \footnote{
Note that in \cite{CLEO2} 
a different version of series parameterization is used
to fit the shape, hence we do not directly compare the 
fitted parameters.}.

Furthermore, we calculate the total semileptonic widths 
divided by the square of CKM parameters from
\be
\frac{\Gamma(D^0\to \pi^-\ell^+ \nu_\ell)}{
|V_{cd}|^2}=
\frac{G_F^2}{24\pi^3}
\int\limits_{0}^{(m_D-m_{\pi})^2} \!\!\!\!dq^2 
\bigg[\left(\frac{m_D^2+m_\pi^2-q^2}{2m_D}
\right)^2-m_\pi^2\bigg]^{\frac{3}{2}}
|f_{D\pi}^+(q^2)|^2\,,
\label{eq:sl}
\ee
(at $m_l=0$) and the analogous formula for 
$\Gamma(D^0\to K^-\ell^+ \nu_\ell)/|V_{cs}|^2$,
using the predicted shape of the form factors
(with the series parameterization)
and their normalization at $q^2=0$.
 Again, in the case of $D\to \pi$ 
a better accuracy is achieved by normalizing with 
the product $f_Df^+_{D\pi}(0)$ calculated from LCSR. 
To this end, multiplying both sides  of (\ref{eq:sl}) by $f_D^2$ 
(in the isospin limit) we replace this factor on l.h.s. 
by the leptonic width,  using
\be
\Gamma(D^+\to \ell^+ \nu_\ell)=\frac{G_F^2}{8\pi}|V_{cd}|^2 f_D^2 m_\ell^2 m_D\left(1-\frac{m_\ell^2}{m_D^2}\right)^2\,,
\label{eq:lept}
\ee
and obtain the following prediction: 
\be
\frac{\Gamma(D^0\to \pi^-\ell^+ \nu_\ell)
\Gamma(D^{\pm}\to \ell^\pm \nu_\ell)}{|V_{cd}|^4}=
(4.7^{+1.4}_{-0.9})\cdot 10^{-28}\,{\mbox GeV}^2\,.
\ee
Employing the experimental numbers 
for the branching fractions from the latest CLEO
measurements: 
$BR(D^0\to \pi^-e^+ \nu_\ell)= 0.288\pm 0.008\pm 0.003 \%$
\cite{CLEO2}, 
$BR(D^+\to  \mu^+\nu_\mu)= (3.82\pm 0.32\pm 0.09)\times
10^{-4}$
\cite{CLEOfD}, and using 
$\tau_{D^\pm}=(1.040\pm 0.007)\,ps$,
$\tau_{D^0}=(0.4101\pm 0.0015)\,ps$ \cite{PDG}, 
 we obtain 
\be\label{eq:vcd_width}
|V_{cd}|= 0.221 \pm 0.002 \pm 0.005\,^{+0.017}_{-0.011}\,,
\ee
with the errors originating from the semileptonic and 
leptonic branching fractions and theoretical uncertainty,
respectively.
This determination is somewhat independent 
of (\ref{eq:vcd}) because it  involves also 
the measured shape of  $f^+(q^2)|V_{cd}|$.
For the ratio of total semileptonic widths we obtain:
\be 
\frac{|V_{cs}|^2
}{|V_{cd}|^2}\frac{\Gamma(D^0\to \pi^-\ell^+ \nu_\ell)}
{\Gamma(D^0\to K^-\ell^+ \nu_\ell)}= 1.65 \pm 0.2\,.
\label{eq:ratio2}
\ee
Substituting the CLEO results
for the branching fractions of these channels: the one quoted above 
and $BR(D^0\to K^-e^+ \nu_\ell)= 3.50\pm 0.03\pm 0.04 \%$ \cite{CLEO2}, 
yields 
\be
\frac{|V_{cd}|}{|V_{cs}|}= 0.223 \pm 0.003 \pm 0.002 \pm 0.015\,.
\ee
where the errors are from semileptonic $D\to\pi$, $D\to K$ 
branching fractions and the LCSR result (\ref{eq:ratio2}), 
respectively. To shorten this discussion, we do not present 
here a comparison 
with the data of other experiments on charm semileptonic decays 
\cite{FOCUS,BELLE,BABAR}, having in mind their 
general agreement with the CLEO data. 

\begin{figure}[t!]
\centering
 \includegraphics[scale=0.27]{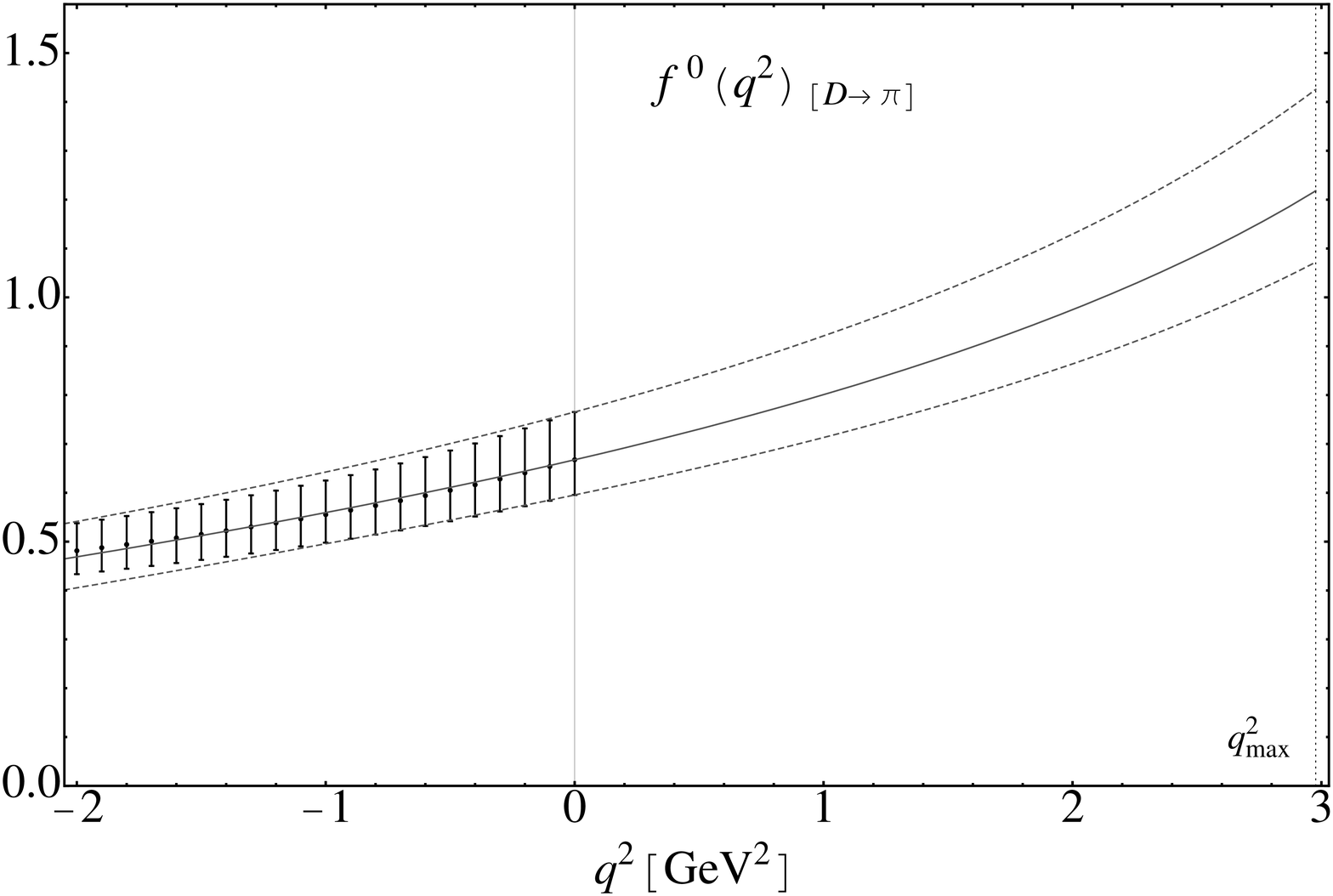}\\[3mm]
 \includegraphics[scale=0.27]{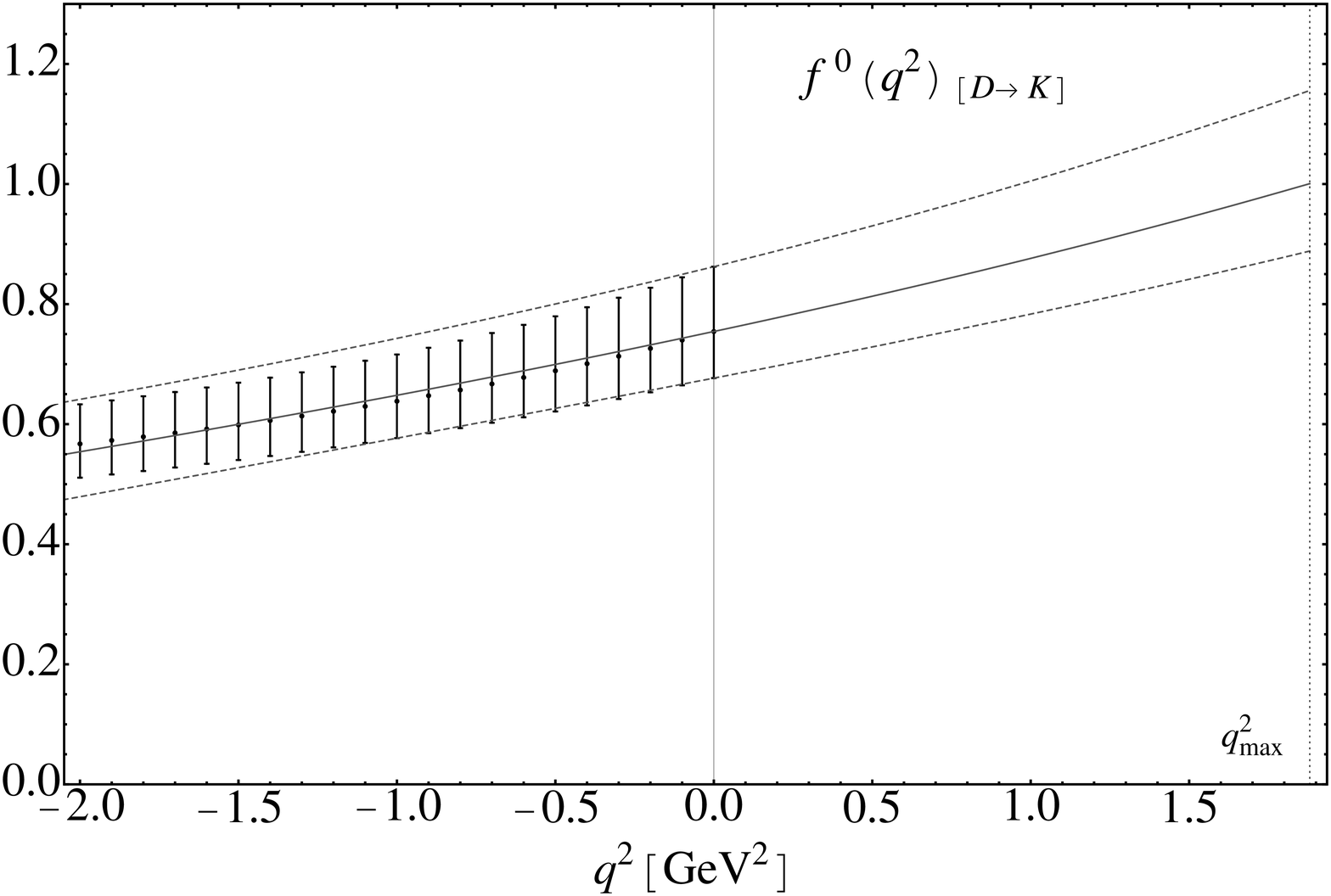}
 \caption{\it
Scalar form factors $f^0_{D\pi(K)}$ 
obtained from fitting the LCSR result 
at negative $q^2$ to the series parameterization.
Notations the same as in Fig.~\ref{fig:ffvsCLEO}.
}\label{fig:f0}
\end{figure}

Concluding this section, we turn to 
the scalar form factor $f^0_{D\pi(K)}(q^2)$ 
which is obtained, substituting the LCSR  results for 
$f^+_{D\pi(K)}(q^2)$ and 
$[f^+_{D\pi(K)}(q^2)+f^-_{D\pi(K)}(q^2)]$ in (\ref{eq:f0}).
We fit the scalar form factors calculated at negative 
$q^2$ to the series parameterization of the type (\ref{eq:BCLfp0})
(without the $D^*_{(s)}$-pole factor 
which is irrelevant in this case). The results for $N=2$ are
\be
b_1^{f^0,D\pi} =-2.6^{+0.3}_{-0.5}, ~~b_1^{f^0,DK} =-3.3^{+0.6}_{-0.8}\,.
\label{eq:fitf0zparam}
\ee
Going to $N\geq 3$ demands 
dedicated unitarity bounds in the scalar heavy-light channel
which, to our knowledge have not been derived and 
are beyond our scope.

The predicted scalar form factors 
are plotted in Fig.~{\ref{fig:f0}}. 

Our results are in agreement 
with $f^0_{D\pi}$ and $f^0_{DK}$ 
presented in \cite{Aubin05} in a form of BK parameterization 
\be
f^0_{D\pi(K)}(q^2)=\frac{1}{1-q^2/(\beta_{D\pi(K)}m_{D^*_{(s)}}^2)}\,,
\label{eq:BKfitf0}
\ee 
with 
$ \beta_{D\pi}=1.41\pm 0.06\pm 0.07$, $\beta_{DK}=1.31\pm 0.07\pm 0.13$.
Our fit to (\ref{eq:BKfitf0}) yields:
$ \beta_{D\pi}=1.25\pm 0.2, ~~ \beta_{DK}=1.3^{+0.4}_{-0.3}  $,  
with larger uncertainties.

Finally, we extrapolate  the scalar form factors 
to the unphysical point $q^2=m_D^2$, located 
slightly above $t_-$, and obtain
\ba 
f^0_{D\pi} (m_D^2) = 1.40^{+0.21}_{-0.14}\,,~~~
f^0_{DK}(m_D^2) = 1.29^{+0.23}_{-0.16}\,.
\label{eq:f0_results} 
\ea
These  results can be compared to the approximate relation 
\be
\lim_{q^2\to m_{D}^2}f^0_{D\pi(K)}(q^2)=f_D/f_{\pi(K)}=
1.58 \pm 0.07~~(1.32 \pm 0.06)\,, 
\label{CT}
\ee
derived from the current algebra combined with the soft pion (kaon)
limit \cite{relf0}, where we used the 
measured values \cite{PDG,CLEOfD} of the decay constants.

\section{Summary}

In this paper we returned to the calculation
of $D\to \pi$ and $D\to K$ form factors
from LCSR's.  Several  improvements have been  implemented, 
including the use of $\overline{MS}$ $c$-quark mass, 
and updated parameters of pion and kaon DA's.

The main advancement  is in the  phenomenological direction.
Employing the accurate measurement of the $D$-meson
decay constant, we effectively decreased 
the theoretical uncertainty of $|V_{cd}|$ determination
from the $D\to \pi l \nu_l $ decay distribution, 
using the LCSR prediction for the product of the $D\to\pi$ 
form factor and $f_D$. The uncertainty in the determination
of $|V_{cs}|$ is also  reduced, due to a better knowledge
of the $s$-quark mass and various $SU(3)_{fl}$ violating effects
in the kaon DA's. Our results for $|V_{cd}|$ and $ |V_{cs}|$ 
are in agreement  with lattice QCD determinations.
 
A new element presented  in this paper 
is the prediction of the $D\to \pi,K$ form factors in the whole 
semileptonic region, combining LCSR calculation 
with the analyticity of the form factors. The latter property 
is cast in the form of conformal mapping and  
series parameterization,
in the version recently suggested in \cite{BCL}.
The form factor shapes obtained from this combined 
procedure are in a good agreement with the latest
experimental measurements of the semileptonic 
charm decay distributions by CLEO collaboration.
Our analysis based on the conformal mapping 
can be further refined, 
by using more terms in the power series and 
implementing the constraints from the dedicated unitarity bounds.
Applications to other hadronic form factors calculated
from LCSR are also possible.

Another interesting task which will 
be studied elsewhere, is the comparison of $B\to \pi$ 
and $D\to \pi$ form factors, calculated at two different
finite quark masses, with the combined heavy-quark mass 
and large recoil limit. Such a comparison will allow one 
to quantify 
the deviations from the symmetry relations
in this limit, as well as the proportion of hard-scattering  
and soft-overlap mechanisms in the heavy-light form factors.

LCSR's provide analytical, but essentially approximate
expressions for the hadronic form factors. 
The accuracy of light-cone OPE is limited, 
due to finite amount of terms in the twist expansion 
and uncertainties in the parameters of pion and kaon DA's. 
A further improvement of OPE is possible,
e.g., if the subleading twist-5 terms are calculated.
For that one needs a separate study of pion and kaon twist-5 DA's.
The gluon radiative corrections to 
the subleading twist-4 and three-particle terms
represent a technically difficult task, but we  expect 
these corrections to be very small. 
A further limitation of the accuracy is caused by 
the quark-hadron duality 
approximation used to model the hadronic spectral density
in LCSR. The resulting uncertainty is difficult
to estimate, still it is effectively  minimized in the sum rules. 

The comparison of our predictions 
for charm semileptonic decays with experiment and lattice QCD 
ensures optimism and provides an 
additional test for the important applications of the LCSR method, 
such as  the $|V_{ub}|$ determination from exclusive semileptonic 
$B\to \pi$  decays.

\vspace{1cm}

\noindent {\bf Acknowledgements}\\[1mm]

This work is supported by the Deutsche Forschungsgemeinschaft
under the  contract No. KH205/1-2.
Work of N.O. is supported by EU Contract No.MRTN-CT-2006-035482, 
``FLAVIAnet''. A.K. also acknowledges the travel support of 
``FLAVIAnet''  during his visit to Orsay.
\clearpage

\appendix

\appendix

\section*{Appendix}

\section{Summary of the $K$-meson distribution amplitudes}
\label{app:inputDA}

Here we present the definitions and the 
expressions of the two- and three-particle 
$K$-meson DA's of twist 2,3 and 4 used in LCSR.
The corresponding formulae for pion DA's  are obtained 
by replacing everywhere $K\to \pi$, $s\to d$,  $m_K\to m_\pi$, 
$m_s\to m_{d}\to 0$ and $\mu_K\to \mu_\pi$. 

\subsection{Definitions}

The K-meson two-particle DA's
are defined from the following bilocal matrix element:
\ba
&& \langle K^-(p)|\bar{s}_\omega^i(x_1) 
u^j_\xi(x_2)|0\rangle_{x^2\to 0}
  = \frac{i\delta^{ij}}{12}f_K 
\int_0^1 du~e^{iu p\cdot x_1 +i\bar{u}p\cdot x_2}
\Bigg ( [\DS p \gamma_5]_{\xi\omega} \varphi_K(u)
\nonumber\\
&&\qquad\qquad -[\gamma_5]_{\xi\omega}\mu_K\phi^p_{3K}(u)
+\frac 16[\sigma_{\beta\tau}\gamma_5]_{\xi\omega}p_\beta(x_1-x_2)_\tau \mu_K\phi^\sigma_{3K}(u)
\nonumber\\
&&\qquad\qquad  +\frac1{16}[\DS p \gamma_5]_{\xi\omega}(x_1-x_2)^2\phi_{4K}(u)
-\frac{i}{2} [(\DS x_1-\DS x_2) \gamma_5]_{\xi\omega}
\int\limits_0^u\psi_{4K}(v)dv 
\Bigg)
\,,
\label{eq:2part}
\ea
where the variable $u$ ($\bar u = 1-u$) is the fraction of the meson 
momentum carried by the $s$-quark (light antiquark). This 
decomposition contains the twist-2 DA $\varphi_K(u)$, twist-3 DA's 
$\phi^p_{3K}(u)$ and $\phi^\sigma_{3K}(u)$, and twist-4 
DA's $\phi_{4K}(u)$ and $\psi_{4K}(u)$.
The  definition of each separate DA is easily obtained, 
multiplying both parts of the above equation by the 
corresponding combinations of $\gamma$ matrices and 
taking Dirac and color traces.

In the same way, 
the three-particle DA's are defined via the matrix element: 
\ba
&& \langle K^-(p)|\bar{s}_\omega^i(x_1) g_sG_{\mu\nu}^a(x_3)
u^j_\xi(x_2)|0
\rangle_{x^2\to 0}
=\frac{\lambda^a_{ji}}{32}\int {\cal D}\alpha_ie^{ip(\alpha_1 x_1+\alpha_2 x_2+\alpha_3x_3)}
\nonumber
\\
&& \times\Bigg[if_{3K}(\sigma_{\lambda\rho}\gamma_5)_{\xi
\omega}(p_\mu p_\lambda g_{\nu\rho}-
p_\nu p_\lambda g_{\mu\rho})\Phi_{3K}(\alpha_i)
\nonumber
\\
&& -f_K(\gamma_\lambda\gamma_5)_{\xi\omega}\Big\{(p_\nu 
g_{\mu\lambda}-p_\mu g_{\nu\lambda})\Psi_{4K}(\alpha_i)
+\frac{p_\lambda(p_\mu x_\nu-p_\nu x_\mu)}{(p\cdot x)}
\left(\Phi_{4K}(\alpha_i)+\Psi_{4K}(\alpha_i)\right)
\Big\}
\nonumber
\\
&& -\frac{if_K}2\epsilon_{\mu\nu\delta\rho}(\gamma_\lambda)_{\xi\omega}
\Big\{(p^\rho g^{\delta\lambda}-p^\delta g^{\rho\lambda})
\widetilde{\Psi}_{4K}(\alpha_i)+  
\frac{p_\lambda(p^\delta x^\rho-p^\rho x^\delta)}{(p\cdot x)}
\left(\widetilde{\Phi}_{4K}(\alpha_i)+
\widetilde{\Psi}_{4K}(\alpha_i)\right)\Big\}\Bigg]\,.
\nonumber 
\\
\label{eq:3part}
\ea
where $\Phi_{3K}(\alpha_i)$ is of twist-3
and the other four DA's are of twist-4.

In addition, in~\cite{BBL} one more three-particle twist-4 
DA $\Xi_{4K}(\alpha_i)$ is taken into consideration, 
originating from the operator which contains a covariant derivative 
of the gluon field:

\be
\left\langle K(p)\left \vert \bar{s}(x_1)
\gamma_{\mu}  \gamma_{5} gD^{\alpha}G_{\alpha\beta}(x_3)q(x_2)
\right\vert 0\right\rangle
=
-if_{K}p_{\mu}p_{\beta}
\int \!{\cal D}\alpha_i\,\,
e^{ip(\alpha_1 x_1+\alpha_2 x_2+\alpha_3x_3)}\,\Xi_{4K}(\alpha_i).
\label{eq:3part_Xi}
\ee

The three-particle DA's depend on momentum fraction variables 
\linebreak $\alpha_i = \{\alpha_1,\alpha_2,\alpha_3\}$ and the 
integration measure is ${\cal D}\alpha_i=
d\alpha_1 d\alpha_2 d\alpha_3 \delta(1-\alpha_1-\alpha_2-\alpha_3)$.\\

For the total antisymmetric tensor we use the convention $\epsilon^{0123}=-1$, which corresponds to \mbox{$Tr\{\gamma^5\gamma^\mu\gamma^\nu\gamma^\alpha\gamma^\beta\}=4\textit{i}\epsilon^{\mu\nu\alpha\beta}$}.

In the following, we list the expressions for all DA's
entering (\ref{eq:2part}), (\ref{eq:3part}) and (\ref{eq:3part_Xi}),
based on NLO in conformal expansion and operator identities
and updated in \cite{BBL}.

\subsection{Twist-2 distribution amplitudes}

The twist-2 $\varphi_K(u)$ distribution amplitude is expanded in 
a series of Gegenbauer polynomials:
\be\label{eq:confexp}
\varphi_K(u,\mu) = 6 u (1-u) \left( 1 + \sum\limits_{n=1,2,\cdots}
  a^K_{n}(\mu) C_{n}^{3/2}(2u-1)\right),
\ee
where only the first two coefficients (Gegenbauer moments)
$a^K_{1}(\mu)$ and  $a^K_{2}(\mu)$ 
are retained and LO scale-dependence is taken into account.
The formulae for the scale-dependence of these and other 
relevant DA parameters can be found, e.g., in \cite{BBL}.

\subsection{Twist-3 distribution amplitudes}

In the same approximation, the twist-3 DA's 
are described by $\mu_K$ and three additional 
parameters $f_{3K},\omega_{3K},\lambda_{3K}$. Their definitions 
in terms of hadronic matrix elements of local operators
are given in \cite{BBL}. We also use the 
short-hand notation 
$$
\eta_{3K} = \frac{f_{3K}}{f_K \mu_K}\,.
$$
In our calculation we neglect the $u,d$  quark masses, hence the 
expressions presented here are somewhat simpler than the 
original ones in \cite{BBL}.
In particular, in the adopted approximation the parameters 
$\rho^K_+,\rho^K_-$ introduced in~\cite{BBL} are equal:
\be
\rho^K_+ = \rho^K_- \equiv \rho^K = \frac{m_s}{\mu_K}\,.
\qquad 
\ee
The twist-3 kaon DA's used in our calculation are:
\ba
\phi^p_{3K}(u) &=& 1 + 3 \rho^K 
\big(1-3 a^K_1 +6 a^K_2\big)(1+\ln u)
\nonumber\\
&-&\frac{\rho^K}{2}\big(3 - 27 a^K_1  + 54 a^K_2\big)C_1^{1/2}(2u-1)
\nonumber\\
&+& 3  \bigg( 10 \eta_{3K} - \rho^K\big(a^K_1 - 5 a^K_2\big) \bigg)C_2^{1/2}(2u-1) 
\nonumber\\
&+& \bigg( 10 \eta_{3K} \lambda_{3K}
- \frac{9}{2} \rho^K a^K_2\bigg)C_3^{1/2}(2u-1)
- 3\eta_{3K} \omega_{3K} C_4^{1/2}(2u-1)\,, 
\label{eq:phip}
\ea
\ba
\phi^\sigma_{3K}(u)&=&   
6 u\bar u \Bigg\{ 1 + \frac{\rho^K}{2} \big(3 - 15 a^K_1 + 30 a^K_2\big) 
\nonumber\\
&+& \rho^K \bigg( 3 a^K_1 - \frac{15}{2} a^K_2\bigg) C_1^{3/2}(2u-1)
\nonumber\\
&+&\frac{1}{2}\bigg(\eta_{3K} (10 -\omega_{3K})  +
    3 \rho^K a^K_2 \bigg) C_2^{3/2}(2u-1) + 
\eta_{3K} \lambda_{3K}C_3^{3/2}(2u-1)
\nonumber\\
&+&  3 \rho^K \big(1-3 a^K_1+6 a^K_2\big) \ln u \Bigg \}\,,
\label{eq:phisigma}\\
{\Phi}_{3K}(\alpha_i)& =& 360 \alpha_1\alpha_2\alpha_3^2
\left\{ 1 + \lambda_{3K} (\alpha_1-\alpha_2) +
\omega_{3K}\, \frac{1}{2}\left( 7\alpha_3-3\right)\right\}.
\label{eq:turk}
\ea

\subsection{Twist-4 distribution amplitudes}

As explained in detail in \cite{BBL}, the twist-4 DA's 
are described  by 13 parameters of the conformal expansion. 
They  are expressed via three nonperturbative parameters 
$\delta_K^2, \omega_{4K}, \kappa_{4K}$ and in addition fixed
by the  renormalon model of twist-4 DA's. 
Here we give the expressions for the twist-4 kaon DA's, 
where the above mentioned relations are already substituted 
and the same approximation as for twist-3 DA's is adopted. 

We rederived the expressions (4.27) and (4.28) 
in \cite{BBL} for the twist-4 two-particle DA's defined 
in (4.26) and using the operator relations given there in (A1),(A2).
We found that both (4.27) and (4.28) should be corrected
by replacing $\psi_{4;K}(u) \to \psi_{4;K}(u)+m_K^2\phi_{2;K}$
(in the notations of \cite{BBL}).
In fact, our version of $\psi_{4;K}(u)$ agrees with the 
function $B(u)$ introduced in \cite{Ball_DA}.
Moreover, we restore the correct 
normalization $\int_0^1\psi_{4;K}(u)du =0$.

We use the following expressions for two-particle twist-4 DA's:
\ba
\psi_{4K}(u) = \psi_{4K}^{T4}(u) + \psi_{4K}^{WW}(u)\,,
\label{eq:T4psi}
\ea
where
\vspace{-1mm}
\ba
\psi_{4K}^{T4}(u)& =&  \delta_K^2\Bigg\{\frac{20}{3}\, C_2^{1/2}(2u-1) + \frac{49}{2} a_1^K C_3^{1/2}(2u-1)\Bigg\}\,,\label{eq:T4psiCE}
\label{eq:psi4KT4}
\ea
and (the corrected version)
\vspace{-1mm}
\ba
\psi_{4K}^{WW}(u)& =& m_K^2\Bigg\{
\bigg[6\rho^K (1- 3 a_1^K+6 a_2^K)\bigg] C_0^{1/2}(2u-1)
\nonumber\\
&-& \bigg[ \frac{18}{5}\,a_1^K +3\rho^K (1- 9 a_1^K+18 a_2^K) + 
12 \kappa_{4K} \bigg]C_1^{1/2}(2u-1)
\nonumber\\
&+& \bigg[2- 6\rho^K (a_1^K-5 a_2^K)+ 60\eta_{3K}\bigg]C_2^{1/2}(2u-1)
\nonumber\\
&+&\big[\frac{18}{5}\,a_1^K -9 \rho^K a_2^K + \frac{16}{3}\,\kappa_{4K}
+ 20\eta_{3K}\lambda_{3K}\bigg]C_3^{1/2}(2u-1)
\nonumber\\
&+&\bigg[\frac{9}{4}\, a_2^K - 6\eta_{3K}\omega_{3K}\bigg]
C_4^{1/2}(2u-1)\Bigg\}
+6 m_s^2(1-3a_1^K+6a_2^K)\ln u\,,
\label{eq:psi4KWW}
\ea

\be
\phi_{4K}(u) = \phi_{4K}^{T4}(u) + \phi_{4K}^{WW}(u)\,,
\label{eq:T4phi}
\ee
where
\ba
\phi_{4K}^{T4}(u) & = & \delta_K^2\bigg\{\left(\frac{200}{3} + 
196 (2u-1) a_1^K\right)u^2\bar u^2\nonumber\\
&+& 21 \omega_{4K} \bigg( u\bar u (2+13 u\bar u) + \big[ 2 u^3(6
  u^2-15u+10)\ln u\big] + \big[u\leftrightarrow\bar u\big]\bigg)
\nonumber\\
&-&14 a_1^K \bigg( u\bar u (2u-1) (2-3 u\bar u) - 
\big[ 2 u^3(u-2)\ln u\big] + 
\big[u\leftrightarrow \bar u\big]\bigg)\bigg\},
\label{eq:T4phiCE}\\
\phi_{4K}^{WW}(u) & = & m_K^2\bigg\{\frac{16}{3}
\kappa_{4K}\bigg(u\bar u (2u-1) (1-2u\bar u) +
\big[5(u-2)u^3 \ln u\big] - \big[u\leftrightarrow\bar u
\big]\bigg)
\nonumber\\
&+&4\eta_{3K}u\bar u \bigg(60\bar u + 10\lambda_{3K} \big[(2u-1)
(1-u\bar u)-(1-5u\bar u)\big] 
\nonumber\\
&-& \omega_{3K} \big[ 3-21 u\bar u +28 u^2\bar u^2 +
3 (2u-1) (1-7u\bar u)\big]\bigg)
\nonumber\\
&-&\frac{36}{5}a_2^K \bigg (
\frac{1}{4} u\bar u (4-9u\bar u +110 u^2\bar u^2) +
 [u^3(10-15u+6u^2)\ln u] +
[u\leftrightarrow\bar u]\bigg)
\nonumber\\
&+&4u \bar u\,(1+3u\bar u)\left(1+\frac{9}{5}(2u-1) a_1^K 
\right)\bigg\}\,.
\ea

\begin{table}[t!]
\caption{\it Parameters of the pion and kaon DA's 
(normalized at 1 GeV): $a^\pi_2, a^\pi_4$ are fitted in
\cite{DKMMO}, $a_1^K$ from \cite{CKP}, all others from \cite{BBL}. 
$\kappa_4^K$ is calculated from (\ref{eq:BL04}).} 
\vspace{-0.1cm}
\begin{center}
{\small
\begin{tabular}{|c|c||c|c|}
\hline
$a_1^\pi$ & 0 &  $a_1^K$ & $0.10 \pm 0.04$\\
\hline
 $a_2^\pi$ & $0.16 \pm 0.01$ & $a_2^K$ & $0.25 \pm 0.15$ \\
\hline
 $a_4^\pi$ & $0.04 \pm 0.01$ &  $a_4^K$ & $0$ \\
\hline
 $a_{>4}^\pi$& $0$ & $a_{>4}^K$& $0$ \\
\hline
 $f_3^\pi$& $(0.0045 \pm 0.0015)\,\mbox{GeV}^2$ &  $f_3^K$ & $(0.0045 \pm 0.0015)\,\mbox{
GeV}^2$\\
\hline
 $\omega_3^\pi$& $-1.5 \pm 0.7$ & $\omega_3^K$& $-1.2\pm 0.7$\\
\hline
 $\lambda_3^\pi$& 0 & $\lambda_3^K$& $1.6 \pm 0.4$\\
\hline
 $\omega_4^\pi$& $0.2 \pm 0.1$ &  $\omega_4^K$& $0.2 \pm 0.1$\\
\hline
 $\delta_\pi^2$ & $(0.18 \pm 0.06)\,\mbox{GeV}^2$ & $\delta_K^2$ & $(0.2 \pm 0.06)\,\mbox{GeV}^2$\\
\hline
$\kappa_{4\pi}$ &0 & $\kappa_{4K}$ & $-0.12 \pm 0.01$\\
\hline
\end{tabular}
}
\end{center}
\label{tab:parampiK}
\end{table}

The twist-4 three-particle DAs have 
the following expressions:
\ba
\Phi_{4K}(\alpha_i) & = & 120 \alpha_1 \alpha_2 \alpha_3 
\Bigg\{\delta_K^2\Big[\frac{21}{8} (\alpha_1-\alpha_2)\omega_{4K} 
+\frac{7}{20} a_1^K (1-3 \alpha_3)\Big]
\nonumber\\
&+&
m_K^2\Big[ -\frac{9}{20} (\alpha_1-\alpha_2)a_2^K 
+\frac{1}{3}\kappa_{4K}\Big]\Bigg\}\,,
\label{eq:Phi4K}
\ea

\ba
\widetilde\Phi_{4K}(\alpha_i) & = -& 120 \alpha_1 \alpha_2 \alpha_3 
\delta_K^2 \Big\{\frac13 +\frac74\,a_1^K (\alpha_1-\alpha_2)+
\frac{21}8\omega_{4K}(1-3 \alpha_3)
\Big\},
\label{eq:phi}\\
%
{\Psi}_{4K}(\alpha_i) & = & 30 \alpha_3^2
\Bigg\{\delta_K^2\bigg[\frac13(\alpha_1-\alpha_2)+\frac{7}{10}
 a_1^K\big[-\alpha_3(1-\alpha_3)+3(\alpha_1-\alpha_2)^2 \big]
\nonumber\\
&+&
\frac{21}4\omega_{4K}(\alpha_1-\alpha_2)(1-2\alpha_3)\bigg]
\nonumber\\
&+& m_K^2(1-\alpha_3)\bigg[
\frac{9}{40}(\alpha_1-\alpha_2)-\frac13\kappa_{4K}\bigg]\,,
\label{eq:Psi4K}
\ea
\ba
{\widetilde\Psi}_{4K}(\alpha_i) & = &30 \alpha_3^2
\Bigg\{\delta_K^2\bigg[\frac13(1-\alpha_3)-
\frac{7}{10} a_1^K(\alpha_1-\alpha_2)(4\alpha_3-3)
\nonumber\\
&+&
\frac{21}4 \omega_{4K}(1-\alpha_3)(1-2\alpha_3) \bigg]
\nonumber\\
&+&
m_K^2\bigg[\frac{9}{40}a_2^K(\alpha_1^2-4 \alpha_1 \alpha_2+\alpha_2^2)-
\frac{1}3(\alpha_1-\alpha_2)\kappa_{4K}\bigg]\Bigg\}\,,
\label{eq:Psitilda}
\ea
\be
  \Xi_{4K}(\alpha_i) = 840 \alpha_1\alpha_2\alpha_3^3\,\Xi^K_0,
\label{Xi1}
\ee

Using the equations of motion, the parameter $\kappa_{4K}$ can be expressed 
via $a_1^K$ and the quark mass:
\be
\kappa_{4K} = -\frac{1}{8} - \frac{9}{40}\,
a_1^K + \frac{m_s}{2\mu_K},
\label{eq:BL04}
\ee
and the parameter entering $\Xi_{4K}$ is taken from 
the renormalon model \cite{BBL}:
\be
\Xi_0^K = \frac{1}{5}\delta_K^2 a_1^K
\ee

The numerical values for all parameters entering the pion- and kaon- DA's 
are collected in Table~\ref{tab:parampiK}.

\section{Contributions to LCSR}

Here we present the separate contributions 
to LCSR's for the form factors $f^+_{DK}$ and $(f^+_{DK}+f^-_{DK})$
including LO twist-2,-3 and -4 terms.
The corresponding contributions for $D\to\pi$ form factors 
are obtained by replacing: the kaon DA's by the pion DA's, 
$m_K \to m_\pi \simeq 0$ and $m_s \to m_d \simeq 0$.

\ba
&&F^{K,2}_0(q^2,M^2,s_0^D)= m_c^2f_K\int\limits_{u_0}^1 
\frac{du}{u}\,
e^{-\frac{m_c^2-q^2\bar{u} + m_K^2 u \bar{u}}{u M^2}} \varphi_K(u)\,,
\label{eq:fplusDpiLCSRcontribTw2}
\ea
\ba
&&F^{K,3}_0(q^2,M^2,s_0^D)= 
m_c^2f_K\int\limits_{u_0}^1 du\,
e^{-\frac{m_c^2-q^2\bar{u} + m_K^2 u \bar{u}}{u M^2}}
\Bigg\{
\frac{\mu_K}{m_c}\Bigg[ \phi_{3K}^p(u)
\nonumber
\\
&& \qquad 
+\frac{1}{3}\Bigg( \frac{1}{u}
-\frac{m_c^2 +q^2 - u^2 m_K^2}{2(m_c^2 -q^2 + u^2 m_K^2)}\frac{d}{du}
- \frac{2 u m_K^2 m_c^2}{(m_c^2 -q^2 + u^2 m_K^2)^2}\Bigg)
\phi_{3K}^\sigma(u)\Bigg]
\nonumber
\\
&& \qquad 
-\frac{f_{3K}}{m_c f_K} 
\Bigg[
\frac{2}{u}\left(\frac{m_c^2-q^2 - u^2 m_K^2}{m_c^2-q^2 + u^2 m_K^2}\right)\left ( \frac{d}{du} - \frac{2 u m_K^2}{m_c^2-q^2 + u^2 m_K^2} \right )I_{3K}(u)
\nonumber \\
&& \qquad 
+\frac{3m_K^2}{m_c^2-q^2 + u^2 m_K^2}
\left ( \frac{d}{du} - \frac{2 u m_K^2}{m_c^2-q^2 + u^2 m_K^2} \right )
 \bar{I}_{3K}(u) 
\Bigg]
\Bigg\}\,,
\label{eq:fplusDpiLCSRcontribTw3}
\ea

\ba
&& F^{K,4}_0(q^2,M^2,s_0^D)= m_c^2f_K\int\limits_{u_0}^1 du\,
e^{-\frac{m_c^2-q^2\bar{u} + m_K^2 u \bar{u}}{u M^2}}
\Bigg\{
\nonumber
\\
&& \qquad \frac{1}{m_c^2-q^2 + u^2 m_K^2}
\Bigg[
u\psi_{4K}(u)+ \left ( 1 - \frac{2 u^2 m_K^2}{m_c^2-q^2 + u^2 m_K^2} \right )\int\limits_0^u dv \psi_{4K}(v)
\nonumber
\\
&& \qquad
- \frac{m_c^2u}{4(m_c^2-q^2 + u^2 m_K^2)}
\left ( \frac{d^2}{du^2} - \frac{6 u m_K^2}{m_c^2-q^2 + u^2 m_K^2} \frac{d}{du} +
\frac{12 u m_K^4}{(m_c^2-q^2 + u^2 m_K^2)^2} \right ) \phi_{4K}(u)
\nonumber \\
&& \qquad - \left ( \frac{d}{du} - \frac{2 u m_K^2}{m_c^2-q^2 + u^2 m_K^2} 
\right )
\left ( I_{4K}(u)  - \frac{d I_{4K}^{\Xi}(u)}{du}  \right )
\nonumber \\
&& \qquad
-\frac{ 2 u m_K^2 }{m_c^2-q^2 + u^2 m_K^2}
\left ( u\frac{d}{du} + \left (1- \frac{4 u^2 m_K^2}{m_c^2-q^2 + u^2 m_K^2} \right ) \right ) \overline{I}_{4K}(u)
\nonumber
\\
&& \qquad + \frac{ 2 u m_K^2 (m_c^2-q^2 - u^2 m_K^2)}{(m_c^2-q^2 + u^2 m_K^2)^2}
\left ( \frac{d}{du} - \frac{6 u m_K^2}{m_c^2-q^2 + u^2 m_K^2} \right ) \int_u^1 d\xi \overline{I}_{4K}(\xi)
\Bigg]
\Bigg\}
\nonumber \\
&& \qquad +  \frac{m_c^4f_K e^{-\frac{m_c^2}{ M^2}}}{4(m_c^2 -q^2 +  m_K^2)^2}\bigg ( \frac{d \phi_{4K}^{WW}(u)}{du} \bigg )_{u \to 1}  \, ,
\nonumber \\
\label{eq:fplusDpiLCSRcontribTw4}
\ea
where $\bar{u}=1-u$, 
$u_0= \left (q^2 - s_0^D + m_K^2 + \sqrt{(s_0^D-q^2-m_K^2)^2 +
4 m_K^2 ( m_c^2-q^2)} \right )/(2 m_K^2) $,  and the short-hand notations
introduced for the integrals over three-particle DA's are:
\ba
I_{3K}(u)&=&\int\limits_0^u \!d\alpha_1\!\!\!
\int\limits_{(u-\alpha_1)/(1-\alpha_1)}^1\!\!\!\!\! dv \,\,
\, \Phi_{3K}(\alpha_i)
\Bigg|_{\begin{array}{l}
\alpha_2=1-\alpha_1-\alpha_3,\\
\alpha_3=(u-\alpha_1)/v
\end{array}
}
\,,
\label{eq:fplusDpiLCSR3part1}
\ea
\ba
\bar{I}_{3K}(u)&=&u\int\limits_0^u \!d\alpha_1\!\!\!
\int\limits_{(u-\alpha_1)/(1-\alpha_1)}^1\!\!\!\!\! \frac{dv}{v} \,\,
\left(2v - 1\right) \Phi_{3K}(\alpha_i)
\Bigg|_{\begin{array}{l}
\alpha_2=1-\alpha_1-\alpha_3,\\
\alpha_3=(u-\alpha_1)/v
\end{array}
}
\,,
\label{eq:fplusDpiLCSR3part2}
\\
I_{4K}(u)&=&\int\limits_0^u\! d\alpha_1\!\!\!
\int\limits_{(u-\alpha_1)/(1-\alpha_1)}^1\!\!\!\!\! \frac{dv}{v} \,\,
\Bigg[ 2 \Psi_{4K}(\alpha_i)-  \Phi_{4K}(\alpha_i) + 
\nonumber
\\
&&2 \widetilde{\Psi}_{4K}(\alpha_i) - \widetilde{\Phi}_{4K}(\alpha_i)
\Bigg]
\Bigg|_{\begin{array}{l}
\alpha_2=1-\alpha_1-\alpha_3,\\
\alpha_3=(u-\alpha_1)/v
\end{array}
}
\,,
\\
\overline{I}_{4K}(u)&=&\int\limits_0^u\! d\alpha_1\!\!\!
\int\limits_{(u-\alpha_1)/(1-\alpha_1)}^1\!\!\!\!\! \frac{dv}{v} \,\,
\Bigg[
\Psi_{4K}(\alpha_i) + \Phi_{4K}(\alpha_i) 
\nonumber
\\
&&+\widetilde{\Psi}_{4K}(\alpha_i)+ \widetilde{\Phi}_{4K}(\alpha_i)
\Bigg]
\Bigg|_{\begin{array}{l}
\alpha_2=1-\alpha_1-\alpha_3,\\
\alpha_3=(u-\alpha_1)/v
\end{array}
}
\,,
\\
I_{4K}^{\Xi}(u)&=& \int\limits_0^u \! d\alpha_1\!\!\!
\int\limits_{(u-\alpha_1)/(1-\alpha_1)}^1\!\!\!\!\! \frac{dv}{v} \,\,
\Bigg[
v (1-v) \Xi_{4K}(\alpha_i)
\Bigg]
\Bigg|_{\begin{array}{l}
\alpha_2=1-\alpha_1-\alpha_3,\\
\alpha_3=(u-\alpha_1)/v
\end{array}
}
\,.
\label{eq:fplusDpiLCSR3part}
\ea

The LCSR for ($f^+_{DK} + f^-_{DK}$) in LO has the following contributions:
\be
\widetilde{F}^{K,2}_0(q^2,M^2,m_c^2,s_0^D)= 0\,,
\label{eq:fplminDpiLCSRcontribTw2}
\ee
\ba
&& \widetilde{F}^{K,3}_0(q^2,M^2,s_0^D)=
m_c^2 f_K\int\limits_{u_0}^1 du\, e^{-\frac{m_c^2-q^2\bar{u} + m_K^2 u \bar{u}}{uM^2}}
\Bigg\{\frac{\mu_K}{m_c}\Bigg(\frac{\phi_{3K}^p(u)}{u}
+\frac{1}{6u}\frac{d\phi_{3K}^{\sigma}(u)}{du}\Bigg)
\nonumber \\
&& \qquad + 
\left(\frac{f_{3K}}{f_K m_c}\right) \frac{m_K^2}{m_c^2 - q^2 + u^2 m_K^2} \left (\frac{d}{du} - \frac{2 u m_K^2}{m_c^2 - q^2 + u^2 m_K^2} \right )
\widetilde{I}_{3K}(u)\Bigg\}
\,,
\label{eq:fplminDpiLCSRcontribTw3}
\ea
where
\ba
&& \widetilde{I}_{3K}(u)=\int\limits_0^u \!d\alpha_1\!\!\!
\int\limits_{(u-\alpha_1)/(1-\alpha_1)}^1\!\!\!\!\! \frac{dv}{v} \,\,
\left [ (3 - 2v) \right ] \Phi_{3K}(\alpha_i)
\Bigg|_{\begin{array}{l}
\alpha_2=1-\alpha_1-\alpha_3,\\
\alpha_3=(u-\alpha_1)/v
\end{array}
}
\,,
\ea

and 
\ba
&& \widetilde{F}^{K,4}_0(q^2,M^2,s_0^D)=
m_c^2 f_K\int\limits_{u_0}^1 du\, e^{-\frac{m_c^2-q^2\bar{u} + m_K^2 u \bar{u}}{uM^2}}
\Bigg\{
\frac{1}{m_c^2-q^2 + u^2 m_K^2}
\Bigg [ \psi_{4K}(u) 
\nonumber \\
&& \qquad 
- \frac{2 u m_K^2}{m_c^2 - q^2 + u^2 m_K^2} \int_0^u dv \psi_{4K}(v)
+ \frac{2 u m_K^2}{m_c^2 - q^2 + u^2 m_K^2} \bigg ( \frac{d^2}{du^2} 
\nonumber \\
&& \qquad
- \frac{6 u m_K^2}{m_c^2 - q^2 + u^2 m_K^2} \frac{d}{du}
+ \frac{12 u^2 m_K^4}{(m_c^2 - q^2 + u^2 m_K^2)^2} \bigg ) \int_u^1 d\xi \overline{I}_{4K}(\xi)
\Bigg ]
\Bigg\}\,.
\label{eq:fplminDpiLCSRcontribTw4}
\ea

\end{document}